\documentclass[preprint2]{aastex}
\usepackage{epsf,amsmath}
\usepackage{amssymb}
\usepackage{array,tabularx}

\begin{document}

\title{DESCRIPTION OF THE SCENARIO MACHINE}

\author{V.M. Lipunov\thanks{E-mail: lipunov@xray.sai.msu.ru}, K.A. Postnov\thanks{E-mail: pk@.sai.msu.ru},
M.E. Prokhorov\thanks{E-mail: mystery@xray.sai.msu.ru}, A.I. Bogomazov\thanks{E-mail: a78b@yandex.ru}\\
{\small Sternberg astronomical institute, Universitetskij
prospect, 13, 119992, Moscow, Russia} }

\begin{abstract}
We present here an updated description of the ``Scenario Machine''
code. This tool is used to carry out a population synthesis of
binary stars. Previous version of the description can be found at
http://xray.sai.msu.ru/~mystery//articles/review/contents.html;
see also \citep{lipunov1996b,lipunov1996c}.
\end{abstract}


\section{Basic equations and initial distributions}

We use the current scenario of evolution of binary stellar systems
based upon the original ideas that appeared in the papers by
\citet{paczynski1971a,tutukov1973a}, \\ \citet{heuvel1972a} (see
also review by \citet{heuvel1994a}). The scenario for normal star
evolution was joined with the ideas of neutron star evolution (see
pioneer works by \citet{shvartsman1970a, shvartsman1971a,
shvartsman1971b, illarionov1975a,shakura1975a,biskogan1976a}, \\
\citet{lipunov1976a}, \\ \citet{heuvel1977a}). This joint scenario
has allowed to construct a two-dimensional classification of
possible states of binary systems containing NS
(\citet{kornilov1983a,kornilov1983b,lipunov1992a}). According to
this classification, we will distinguish four basic evolutionary
stages for a normal star in a binary system:

\begin{enumerate}
\renewcommand{\labelenumi}{\theenumi{ --- }}
\renewcommand{\theenumi}{\Roman{enumi}}

\item A main sequence (MS) star inside its Roche lobe (RL);

\item A post-MS star inside its RL;

\item A MS or post-MS star filling its RL; the mass is transferred
onto the companion.

\item A helium star left behind the mass-transfer in case II and
III of binary evolution; may be in the form of a hot white dwarf
(for $M\le 2.5 M_{\odot}$), or a non-degenerate helium star (a
Wolf-Rayet-star in case of initial MS mass $> 10 M_{\odot}$).

\end{enumerate}

The evolution of single stars can be represented as a chain of consecutive
stages: I $\to$ II $\to$
compact remnant; the evolution of the most massive single stars probably
looks like I $\to$ II $\to$ IV $\to$ compact remnant. The component of
a binary system can evolve like I $\to$ II $\to$ III $\to$
IV $\to$ compact remnant.

In our calculations we choose the distributions of initial
binary parameters: mass of the primary zero age main sequence
component (ZAMS), $M_1$, the binary mass ratio, $q=M_2/M_1<1$, the orbital
separation $a$. Zero initial eccentricity is assumed.

The distribution of binaries by orbital separations can be taken from
observations \citep{krajcheva1981,abt1983a},

\begin{equation}
\label{axis} \left\{
\begin{array}{l}
f(\log a)=\mbox{const},\\
\mbox{max}(10 R_{\odot}, \mbox{RL} [M_1])< a \le 10^6 R_{\odot};
\end{array}\right.
\end{equation}

\noindent Especially important from the evolutionary point of view
is how different are initial masses of the components (see e.g.
\citet{trimble1983a}). We have parametrized it by a power-law
shape, assuming the primary mass to obey Salpeter's power law:

\begin{equation}
f(M)=M_1^{-2.35},\quad
0.1M_{\odot}<M_1<120M_{\odot},\label{salpeter}
\end{equation}

\begin{equation}
f(q)\sim q^{\alpha_q},\quad q=M_2/M_1<1;
\end{equation}

We should note that some apparently reasonable distributions --
such as both the primary and secondary mass obeying Salpeter's
law, or ``hierarchical'' distributions involving the assumption
that the total binary mass and primary's mass are distributed
according to the Salpeter mass function -- all yield essentially
flat-like distributions by the mass ratio (i.e. with our parameter
$\alpha_q\simeq 0$).

We assume that the neutron star is formed in the core collapse of
the pre-supernova star. Masses of the young neutron stars are
randomly distributed in the range $M_{NS}^{min}$ --
$M_{NS}^{max}$. Initial NS masses are taken to be in the range
$M_{NS}=1.25-1.4M_{\odot}$. The range of initial masses of the
young NSs was based on the masses of the neutron star in the
B1913+16 binary system and of the radio pulsar in the J0737-3039
binary. In the B1913+16 system (the Hulse-Taylor pulsar, radio
pulsar + neutron star) the mass of the neutron star, which is
definitely not accreting matter from the optical donor, is
$M_{NS}=1.3873\pm 0.0006M_{\odot}$
\citep{thorsett1999a,wex2000a,weisberg2003a}. The mass of the
pulsar in the J0737-3039 system (radio pulsar + radio pulsar),
which likewise does not accrete from an optical companion, is
$M_{PSR}=1.250\pm 0.010M_{\odot}$ \citep{lyne2004a}. The mass
range of stars producing neutron stars in the end of their
evolution is assumed to be $M_{n}$ -- $M_{b}$; stars with initial
masses $M>M_{b}$ are assumed to leave behind black holes; $M_{n}$
is taken to be equal to $10M_{\odot}$ in most cases (but in
general this parameter is free). Note that according to some
stellar models, a very massive star ($\approx 50-100M_{\odot}$)
can leave behind a neutron star as a remnant due to very strong
mass loss via powerful stellar wind, so we account for this
possibility in the corresponding models.

We take into account that the collapse of massive star into a
neutron star can be asymmetrical, so that the newborn neutron star
can acquire an additional, presumably randomly oriented in space
kick velocity $w$ (see Section \ref{kick} below for more details).

The magnetic field of rotating compact objects (neutron stars and
white dwarfs) largely define the evolutionary stage of the compact
object in a binary system (\citet{shvartsman1970a, davidson1973a,
illarionov1975a}), so we use the general classification of
magnetic rotating compact objects (see e.g. \citet{lipunov1992a})
in our calculations. The initial magnetic dipole moment of the
newborn neutron star is taken according to the distribution

\begin{equation}
\label{magmoment}
\begin{array}{l}
f(\log \mu) \propto \mbox{const}, \\
10^{28}\le\mu\le 10^{32} \mbox{G cm}^3,
\end{array}
\end{equation}

\noindent The initial rotational period of the newborn neutron
star is assumed to be $\sim 10$ ms.

It is not definitely clear as yet whether the magnetic field of
neutron stars decays or not (see for a comprehensive review
\citet{chanmugam1992a}). We assume that the magnetic fields of
neutron stars decays exponentially on a timescale of $t_{d}$
(usually we take this parameter to be equal to $10^8$,
$5\cdot10^7$ and $10^7$ years). A radio pulsar is assumed to be
``switched on'' until its period $P$ (in seconds) has reached the
``death-line'' defined by the relation $\mu_{30}/P_d^2=0.4$, where
$\mu_{30}$ is the dipole magnetic moment in units of $10^{30}$ G
cm$^3$.

We assumed that magnetic fields of neutron stars decay
exponentially to minimal value $B_{min}=8\cdot 10^7$ G and do not
decay further:

\begin{equation}
 B=\left\{\begin{array}{l}
B_0 \exp(-t/t_d), t < t_d\ln (B_0/B_{min}), \\
B_{min}, t\ge t_d\ln (B_0/B_{min}). \label{field}
\end{array}\right. \end{equation}

\noindent Parameters $B_0$ and $t_d$ in equation (\ref{field}) are
the initial field strength and the field decay time.

We also assume that the mass limit for neutron stars (the
Oppenheimer-Volkoff limit) is $M_{OV}=2.0 M_{\odot}$ (in general,
it is a free parameter in the code; it depends on equation of
state of the material of the neutron star).

The most massive stars are assumed to leave behind black holes
after the collapse, provided that the progenitor mass before the
collapse has a mass $M_{cr}$. The masses of the black holes are
calculated as $M_{bh}=k_{bh}M_{PreSN}$, where the parameter
$k_{bh}=0.0-1.0$, $M_{PreSN}$ is the mass of the pre-supernova
star.

We consider binaries with $M_1\ge 0.8 M_{\odot}$ with a constant
chemical (solar) composition. The process of mass transfer between
the components is treated as conservative when appropriate, that
is the total angular momentum of the binary system is assumed to
be constant. If the accretion rate from one component to another
is sufficiently high (say, the mass transfer occurs on a timescale
few times shorter than the thermal Kelvin-Helmholtz time for the
normal companion) or a compact object is engulfed by a giant
companion, the common envelope stage of binary evolution can begin
\citep{paczynski1976a,heuvel1983a}.

Other cases of non-conservative evolution (for example, stages
with strong stellar wind or those where the loss of binary angular
momentum occurs due to gravitational radiation or magnetic stellar
wind) are treated using the well known prescriptions (see e.g.
\citet{verbunt1981a,rappaport1982a,lipunov1988a}).

\section{Evolutionary scenario for binary stars}

Significant discoveries in the X-ray astronomy made during the last
decades stimulated the astronomers to search for particular
evolutionary ways of obtaining each type of observational
appearance of white dwarfs, neutron stars and black holes, the
vast majority of which harbours in binaries. Taken as a whole,
these ways costitute a general evolutionary scheme, or the
``evolutionary scenario''. We follow the basic ideas about stellar
evolution to describe evolution of binaries both with normal
and compact companions.

To avoid extensive numerical calculations in the statistical
simulations, we treat the continuous evolution of each binary
component as a sequence of a finite number of basic evolutionary
states (for example, main sequence, red supergiant, Wolf-Rayet
star, hot white dwarf, etc.), at which stellar parameters significantly
differ from each other. The evolutionary state of the binary can
thus be determined as a combination of the states of each
component, and is changed once the more rapidly evolving component
goes to the next evolutionary stage.

At each such stage, we assume that the star does not change its
physical parameters (mass, radius, luminosity, the rate and velocity
of stellar wind, etc.) that have effect on the evolution of the
companion (especially in the case of compact magnetized stars).
Every time the faster evolving component passes into the next
stage, we recalculate its parameters. Depending on the
evolutionary stage, the state of the slower evolving star is
changed or can remain unchanged. With some exceptions (such as
the common envelope stage and supernova explosion), states of both
components cannot change simultaneously. Whenever possible we
use analytical approximations for stellar parameters.

Prior to describing the basic evolutionary states of the normal
component, we note that unlike single stars, the evolution of a
binary component is not fully determined by the initial mass and
chemical composition only.  The primary star can fill its Roche
lobe either when it is on the main sequence, or when it has
a (degenerate) helium or carbon-oxygen core.
This determines the rate of mass transfer to the secondary companion
and the type of the remnant left behind. We will follow
\citet{webbink1979a} in treating the first mass exchange modes for
normal binary components, whose scheme accounts for the physical state of
the star in more detail than the simple types of mass exchange (A,
B, C) introduced by \citet{kippenhahn1967a}. We will use both
notations A, B, C and D (for very wide systems with independently
evolving companions) for evolutionary types of binary as a whole,
and Webbink's notations for mass exchange modes for each component
separately [Ia], [Ib], [IIa], [IIb], [IIIa], etc.

\section{Basic evolutionary states of normal stars}

The evolution of a binary system consisting initially of two
zero-age main sequence stars can be considered separately for each
components until a more massive (primary) component fills its Roche
lobe. Then the matter exchange between the stars begins.

The evolutionary states of normal stars will be denoted by Roman
figures (I-IV), whereas those of compact stars will be marked by
capital letters (E, P, A, SA ...). We divide the evolution of a
normal star into four basic stages, which are significant for
binary system evolution and bear a clear physical meaning. We will
implicitly express the mass and radius of the star and the orbital
semi-major axis in solar units ($m\equiv M/M_{\odot}$, $r\equiv
R/R_{\odot}$, $a\equiv A/R_{\odot}$), the time in million years,
the luminosities in units of $10^{38}$ erg s$^{-1}$, the wind
velocities in units of $10^8$ cm s$^{-1}$ and the accretion rates
$\dot M$ onto compact objects in units of $10^{-8}M_{\odot}$
yr$^{-1}$, unless other units are explicitely used.

\subsection{Main sequence stars}

At this stage,
the star is on the zero-age main sequence (ZAMS) and its size is much smaller
than the Roche lobe radius. The time the star spends on the main sequence is
the core hydrogen burning time, $t_H$, which depends
on the stellar mass only \citep{iben1987a}:

\begin{equation}
\left\{
\begin{array}{l}
1.0+0.95\frac{79}{m}, \hspace{36pt} m\ge 79.0, \\
10^{3.9-3.8\log m+\log ^2 m}, \hspace{5pt} 79>m\ge 10, \\
2400 m^{-2.16}, \hspace{40pt} 10>m\ge 2.3, \\
10^4 m^{-3.5}, \hspace{50pt} m<2.3,
\end{array}\right.
\label{th}
\end{equation}

The radius of the ZAMS star is assumed to be

\begin{equation}
\label{rzams} r=\left\{
\begin{array}{l}
10^{0.66 \log m+0.05}, \hspace{15pt} m>1.12, \\
m, \hspace{67pt} m\le 1.2,
\end{array}
\right.
\end{equation}

\noindent and its luminosity is

\begin{equation}
\label{lumin} \log L=\left\{
\begin{array}{l}
-5.032+2.65\log m,\hspace{6pt} (\alpha) \\
-4.253+4.8\log m,\hspace{11pt} (\beta) \\
-4.462+3.8\log m,\hspace{11pt} (\gamma) \\
-3.362+3.0\log m,\hspace{11pt} (\delta) \\
-3.636+2.7\log m,\hspace{11pt} (\epsilon)
\end{array}\right.
\end{equation}

\noindent here we assume the next indication: ($\alpha$), $m<0.6$;
($\beta$), $0.6\le m<1.0$; ($\gamma$), $1.0\le m<10.0$;
($\delta$), $10\le m<48.0$; ($\epsilon$), $m\ge 48.0$.

The initial mass of the primary and the mode of the first mass
exchange (which is determined by the initial orbital period and
masses of the components;  see \citet{webbink1979a}) determine the
mass and the type of the core that will be formed during stage I.
For example, for single stars and primaries in ``type C'' binaries
that fill its Roche lobe having a degenerate core, we use the
expressions

\begin{equation}
\label{mcore} m_c=\left\{
\begin{array}{l}
0.1 m_{max}, \hspace{83pt} (\alpha) \\
0.446+0.106 m_{max}, \hspace{38pt} (\beta) \\
0.24 m^{0.85}_{max}, \hspace{78pt} (\gamma) \\
\mbox{min}(0.36 m_{max}^{0.55}, 0.44 m_{max}^{0.42}), \hspace{6pt} (\delta) \\
0.44 m_{max}^{0.42}, \hspace{78pt} (\epsilon) \\
\approx M_{Ch}, \hspace{89pt} (\zeta) \\
0.1 m^{1.4}_{max}, \hspace{83pt} (\eta)
\end{array}
\right.
\end{equation}

\noindent where $m_{max}$ is the maximum mass the star had during
the preceding evolution. We assume the next indication in this
formula: ($\alpha$), $m_{max}<0.8$; ($\beta$), $0.8\le
m_{max}<2.3$; ($\gamma$), $2.3\le m_{max}<4.0$; ($\delta$),
$4.0\le m_{max}<7.5$; ($\epsilon$), $7.5\le m_{max}<8.8$;
($\zeta$), $8.8\le m_{max}<10.0$; ($\eta$), $m_{max}\ge 10.0$.

A main-sequience star accreting matter during the first mass transfer will
be treated as a rapidly rotating ``Be-star'' with the
stellar wind rate different from what is expected from a single star
of the same mass (see below).

\subsection{Post main-sequence stars}

The star leaves the main sequence and goes toward the red
(super)giant region. The star still does not fill its Roche lobe.
The duration of this stage for a binary component is not any more
a function of the stellar mass only (as in the case of single
stars), but also depends on the initial binary type (A, B, or C)
(see \citet{iben1985a,iben1987a}):

\begin{equation}
\label{tii} t_{II}= \\ \left\{ \begin{array}{l}
0, \hspace{65pt} (\alpha), \\
2t_{KH}, \hspace{48pt} (\beta), \\
6300 m^{-3.2}, \hspace{25pt} (\gamma), \\
t_{He}, \hspace{57pt} (\delta), \end{array} \right.
\end{equation}

\noindent In type A systems, the primaries fill their Roche lobes
when they belong to the main sequence. We assume the next
indication in this formula: ($\alpha$), type A; ($\beta$), type B
excluding mode [IIIA]; ($\gamma$), types C, D and mode [IIIA],
$m_{max}<5$; ($\delta$), types C, D and mode [IIIA], $m_{max}>5$.

The radius of the post-MS star rapidly increases (on the thermal
time scale) and reaches the characteristic giant values. The star
spends the most time of helium burning with such large radius. In
the framework of our approximate description, we take the radius
of the giant star to be equal to the maximum value, which depends
strongly on the mass of its core and is calculated according to
Webbink's mass transfer modes as follows (see also
\citet{iben1985a,iben1987a}):

\begin{equation}
\label{rii} r_{II}=\left\{
\begin{array}{l}
3000 m^4_c, \hspace{66pt} (\alpha) \\
1050 (m_c - 0.5)^{0.68}, \hspace{19pt} (\beta) \\
10 m_c^{0.44}, \hspace{66pt} (\gamma)
\end{array}
\right.
\end{equation}

\noindent We assume the next indication in this formula:
($\alpha$), mode [IIIA] or [IIIB] with He core; ($\beta$), mode
[IIIB] with CO or ONeMg core; ($\gamma$), modes [I] or [II].
Formula (\ref{rii}) depicts stars with mass $\le 10 M_{\odot}$.

This maximum radius can formally exceed the Roche lobe size; in
such cases we put it equal to $0.9 R_L$ during the stage II. The
most sensitive to this crude approximation are binaries with
compact companions, which can lead, for example, to the overestimation of
the number of accreting neutron stars observed as X-ray pulsars.
However, these stages are less important for our
analysis than the stages at which the optical star fills its Roche lobe.
A more detailed treatment of normal star evolution (given,
for example, by \citet{pols1994a}) can reduce such uncertainties.

Luminosities of giants are taken from \citet{jager1980a}:

\begin{equation}
\label{lii} \log l_{II}=\left\{
\begin{array}{l}
\frac{15.92 m_c^6}{(1.0+m_c^4)(2.512+3.162 m_c)}, \hspace{11pt} (\alpha) \\
10^{-4.462+3.8\log m}, \hspace{36pt} (\beta) \\
10^{-3.362+3.0\log m}, \hspace{36pt} (\gamma) \\
10^{-3.636+2.7\log m}, \hspace{36pt} (\delta)
\end{array}
\right.
\end{equation}

\noindent We assume the next indication in this formula:
($\alpha$), $m<23.7$, $m_c<0.7$; ($\beta$), $m<23.7$, $m_c>0.7$;
($\gamma$), $48>m>23.7$; ($\delta$), $m>48$.

Radii of (super)giants are determined by using the effective
temperature $T_{eff}$ and luminosities. Typical effective
temperatures are taken from \cite{allen1973a}:

\begin{equation}
\label{tempii} \log T_{eff}=\left\{
\begin{array}{l}
4.50, \hspace{15pt} (\alpha) \\
3.60, \hspace{15pt} (\beta) \\
3.70, \hspace{15pt} (\gamma)
\end{array}
\right.
\end{equation}

\noindent We assume the next indication in this formula:
($\alpha$), $m>10.0$; ($\beta$), $m<10.0$, type C or D;
($\gamma$), $m<10.0$, type A or B. So, we calculate $R_{II}$ with
mass higher than $10 M_{\odot}$ calculate using formula

\begin{equation}
r_{II}=\exp{\left\{2.3(0.5\log l_{ii}-2\log T_{eff}
+9.7)\right\}}.
\end{equation}

\subsection{Roche lobe overflow}

At this stage the star fills its Roche lobe (RL) and mass transfer
onto the companion occurs. The mass transfer first proceeds on the
thermal time scale (see extensive discussion of this approximaiton in
\citet{heuvel1994a})

\begin{equation}
\label{tkh} t_{KH}\sim 30 m^2 r_*^{-1}(L/L_{\odot})^{-1},
\end{equation}

The common envelope stage (CE) may be formed if the Roche lobe
overflow occurs in the type C system (where the primary has a well developed core)
even for $q<1$; otherwise (for type B systems) we
use the condition $q\le q_{cr}=0.3$ for the CE stage to occur.
Radius of the star at the Roche lobe filling stage is taken to be
that of the equivalent Roche lobe radius \citep{eggleton1983a}:

\begin{equation}
\label{rlr} \frac{R_L}{a(1-e)}=\frac{0.49 q^{2/3}}{0.6 q^{2/3}+\ln
(1+q^{1/3})},
\end{equation}

\noindent Here $a$ is the binary orbital separation and $e$ is the
orbital eccentricity, $q$ is the arbitrary mass ratio.

For $q<0.6$ a more precise approximation can be used:
\begin{equation}
\label{rlr2} \frac{R_L}{a}=0.4622\left(\frac{q}{1+q}\right)^{1/3}.
\end{equation}

A star filling its Roche lobe has quite different boundary
conditions in comparison with single stars. The stellar radius at
this stage is limited by the Roche lobe. If the stellar size
exceeds the Roche lobe, the star can lose matter on a time scale
close to the dynamical one until its radius becomes smaller than
the new Roche lobe size.

Now we consider how the RL-filling star loses matter. Let the star
be in equilibrium and $R_{eq}(M)=R_L(M)$ at the initial moment of
time. When a fraction of mass $\delta m$ is transported to the
companion, the mass ratio $q$ and semi-major axis $a$ of the
binary changes depending on the mass transfer mode assumed (see
below for details). The RL size then becomes equal to
$R_L(M-\delta m)$.

On the other hand, mass loss disturbs equilibrium of the star
(hydrodynamical and thermal). The hydrodynamical equilibrium is
restored on the dynamical timescale $t_d\propto
\left(GM/R^3\right)^{-1/2}$. The stellar radius changes to a value
$R_{ad}(M-\delta m)$ (where ``ad'' means adiabatic), which can be
bigger or smaller than the equilibrium radius $R_{eq}(M-\delta m)$
of the star. The thermal equilibrium establishes on the thermal
time scale $T_{KH}\approx GM^2/R_L$, so after that the stellar
radius relaxes to the equilibrium value $R_{eq}(M-\delta m)$.

Relations $R_{L}(M-\delta m)$, $R_{ad}(M-\delta m)$ and
$R_{eq}(M-\delta m)$ determine the mode of mass transfer during
the RL overflow stage. Following \citet{webbink1985a}, one usually
introduces the logarithmic derivative $\zeta=d \ln R /d \ln M$
($R\propto M^{\zeta}$). It locally fits the real dependence
$R(M)$. Three values of $\zeta$ are relevant:

%\begin{equation}
$$
\zeta_L=\frac{d \ln R_L}{d \ln M},
$$
%\end{equation}
\begin{equation}
\label{ind} \zeta_{ad}=\frac{d \ln R_{ad}}{d \ln M},
\end{equation}
%\begin{equation}
$$
\zeta_{eq}=\frac{d\ln R_{eq}}{d \ln M},
$$
%\end{equation}

\noindent where {\it ``L'', ``ad''} and {\it ``eq''} correspond to
the values of radii discussed above.

Three possible cases are considered depending on $\zeta_i$:

\begin{enumerate}

\item If $\zeta_{ad}<\zeta_L$, the star cannot be inside its RL
regardless of the mass loss rate $(dM/dt<0)$. Such stars lose
their matter in hydrodynamical time scale. Mass loss rate is
limited only by the speed of sound near the inner Lagrangian point
$L_1$. $\zeta_{eq}$ is unimportant because the size of the star
becomes bigger and bigger than $R_L$. The equilibrium is
impossible.

\item $\zeta_{eq}<\zeta_L<\zeta_{ad}$. The star losing mass cannot
be in thermal equilibrium, because otherwise its size would exceed
$R_L$. Nevertheless, in this case $R_{ad}<R_L$. So the
hydrodynamical equilibrium is established. As a result, the star
loses mass on thermal time scale.

%=============================================
\item $\zeta_L<\zeta_{ad},\zeta_{eq}$. In this case the size of
the star losing mass becomes smaller than its RL. The evolutionary
expansion of the star or the binary semi-major axis decrease due
to orbital angular momentum loss via magnetic stellar wind (MSW)
or gravitational radiation (GW) support the permanent contact of
the star with RL. The star then loses mass on a time scale
dictated by ots own evolutionary expansion or on a time scale
corresponding to the orbital angular momentum loss.

\end{enumerate}

For non-degenerate stars, $R_{eq}$ increases monotonically with
$M$. On the other hand, the exponent $\zeta_{ad}$ is determined by
the entropy distribution over the stellar radius which is
different for stars with radiative and convective envelopes. It
can be shown that stars with radiative envelopes should shrink in
response to mass loss, while those with convective envelopes
should expand \footnote{The adiabatic convection in stellar
envelope can be described by the polytropic equation of state
$P\propto \rho^{5/3}$, similar to non-relativistically degenerate
white dwarfs. For such equation of state the mass-radius relation
becomes inverse: $R\propto M^{-1/3}$. For non-degenerate stars
with convective envelopes this relation holds approximately.}.

Therefore, stars with convective envelopes in binaries should
generally have a higher mass loss rate than those with radiative
envelopes under other equal conditions. The next important factor
is the dependence $R_L(M)$. It can found by substituting one of
the relations $a(M)$ (see below) into equation (\ref{rlr}) or into
equation (\ref{rlr2}) and differentiating it with respect to $M$.
For example, assuming the conservative mass exchange when the
total mass and the orbital angular momentum of the system do not
change, one readily gets that the binary semi-major axis decreases
when the mass transfer occurs from the more massive to the less
massive component; $R_L$ decreases of the primary correspondingly.
When the binary mass ratio reaches unity, the semi-major axis
takes on a minimal value. In contrast, if the less massive star
loses its mass conservatively the system expands. In that case the
mass transfer can be stable.

If more massive component with radiative envelope fills its RL,
the mass transfer proceeds on thermal time scale until the masses
of the components become equal \footnote{The mass exchange can
stop earlier if the entire envelope is lost and the stellar core
is stripped (the core has other values $\zeta_{ad}$ and
$\zeta_{eq}$).}. The next stage of the first mass exchange poceeds
in more slower (nuclear) time scale. If the primary has convective
envelope, the mass transfer can proceed much faster on a time
scale intermediate between the thermal and hydrodynamical one, and
probably on the hydrodynamical time scale. In that case the fast
stage of the mass exchange ends when the mass of the donor
decreases to $\sim 0.6$ mass of the secondary companion
\citep{tutukov1973a}. Further mass transfer should proceed  on the
evolutionary time scale.

The process of mass exchange strongly depends on stellar structure
at the moment of the RL overflow. The structure of the star in
turn depends on its age and the initial mass. The moment of the RL
overflow is determined by the mass of the components and by the
initial semi-major axis of the system.
%(the more initial separation of the components, the longer
%duration of evolution before infill of RL). This make it possible
to calculate a diagram in the $M-a$ (or $M-P_{orb}$) plane which
allows us to conclude when the primary in a binary with given
initial parameters fills its RL, what is its structure at that
moment and what type of the first mass exchange is expected. We
use the diagram calculated by \citet{webbink1979a} (see also the
description of modern stellar wind scenarios below).

We distinguish different sub-stages of the RL overflow according
to the characteristic timescales of the mass transfer:

{\bf stage III}:

This is the most frequent case for the first mass transfer phase.
The primary fills its RL and the mass transfer proceeds faster
than evolutionary time scale (if outer layers of the star are
radiative, then it is thermal time scale, if outer layers are
convective, then time scale is shorter, up to hydrodynamical time
scale). This stage comes to the end when mass ratio in the system
changes (``role-to-role transition''), i.e. when the mass of the
donor (mass losing) star is equal to mass of second companion (for
radiative envelopes) or $0.6$ of the mass of the second companion
(for convective envelopes). This stage also stops if the donor
star totally lost its envelope.

{\bf stage IIIe}:

This is the slow (evolutionary driven) phase of mass transfer. We assume it
to occur in short-period binaries of type A. However, it is not
excluded that it may occur after the mass reversal during the first
stage of mass exchange for binaries of type B (see
\citet{heuvel1994a}), e.g. as in wide low-mass X-ray binaries.

{\bf stage IIIs}

This is the specific to super-accreting compact companions substage
of fast mass transfer at which matter escapes from the
secondary companion carrying away its orbital angular momentum.
Its duration is equal to

\begin{equation}
\label{tiiis} t_{IIIs}=t_{KH}\frac{q(1+q)}{2-q-2q^2},
\end{equation}
$$
q=M_a/M_d<0.5,
$$

\noindent here and below subscripts ``a'' and ``d'' refer to the
accreting and donating mass star, respectively. For systems with
small mass ratios, $q<0.5$, this timescale corresponds to an
effective $q$-time shortening of the thermal time for the
RL-overflowing star.

{\bf stages IIIm,g}

At these stages, the mass transfer is controlled by additional losses of orbital
angular momentum $J_{orb}$ caused by magnetic stellar wind
or gravitational wave emission. The characteristic time of the
evolution is defined as $\tau_J=-(J_{orb}/\dot J_{orb})$, and in
the case of MSW is (see \citet{verbunt1981a,iben1987a})

\begin{equation}
\label{taumsw} \tau_{MSW}=4.42\frac{a^5 m_x
\lambda_{MSW}^2}{(m_1+m_2)^2 m_{op}^4},
\end{equation}

\noindent Here $m_{op}$ denotes mass of the low-mass optical star
($0.3<m<1.5$) that is capable of producing an effective magnetic
stellar wind (because only such stars have outer
convective envelopes which are prerequisit
for effecftive MSW), $\lambda$ is a numerical parameter of order
of unity. We have used the mass-radius relation $r\approx m$ for main
sequence stars in deriving this formula. The upper limit of the mass
interval and empirical braking law for main-sequence G-stars are
taken from \citet{skumanich1972a}, the lower limit is determined by
absence of cataclysmic variables with orbital periods
$P_{orb}\approx 3^{h}$
\citep{verbunt1984a,mestel1952a,kawaler1988a,tout1992a,zangrilli1997a}.
We use $\lambda=1$ (see for details \citet{kalogera1998a}).

The time scale of the gravitational wave emission is

\begin{equation}
\label{taugw} \tau_{GW}=124.2\frac{a^4}{m_1 m_2 (m_1 + m_2)}\times
\end{equation}
$$
\times\left(1+\frac{73}{24}e^2+\frac{37}{96}e^4\right)(1-e^2)^{-7/2},
$$

Wether the evolution is governed by MSW or GW is decided by which time scale ($\tau_{MSW}$ or
$\tau_{GW}$) turns out to be the shortest among all appropriate
evolutionary time scales.

{\bf stage IIIwd}

This is a special case where the white dwarf overflows its RL. This stage
is encountered for very short period binaries (like Am CVn stars and
low-mass X-ray binaries like 4U 1820-30)
whose evolution is controlled by GW or MSW. The mass transfer
is calculated using the appropriate time scale (GW or MSW). The radius of
the white dwarf increases with mass $R_{WD}\propto
M^{-1/3}$. This fact, however, does not automatically imply that
the mass transfer is unstable, since the less massive WD fills its RL first.
It can be shown that the mass exchange is always stable in such systems
if the mass ratio  $q< 0.8$. This condition always holds in WD+NS and
WD+BH systems.
%  in WD+WD systems it holds if the initial mass ratio
% was not equal or approximately equal to 1.
WD loses its matter
until its mass decreases to that of a huge Jupiter-like planet ($\sim$ a few
$10^{-3}M_{\odot}$), where the COulomb interaction reverses
the mass-radius relation $R(M)$. Such a planet can approach the secondary companion
of the system due to GW emission until the tidal forces destroy it completely.
The matter of the planet can fall onto the surface of the second
companion ir form a long-living disk around it.
If the second star is a neutron star and its rotation
had been spun up by accretion such that a millisecond radio pulsar appeared, the
planet can be evaporated by relativistic particles emmited by the pulsar
\citep{paczynski1983a,joss1983a,kolb1998a,kalogera1998a}.

The mass loss rate at each of the III-stages is calculated according
to the relation

\begin{equation}
\label{md} \dot M=\Delta M/\tau_i,
\end{equation}

\noindent where $\Delta M$ is the {\it a priori} known mass to be
lost during the mass exchange phase (e.g. $\Delta M=M_1-(M_1+M_2)/2$
in the case of the conservative stage III, or $\Delta
M=M_1-M_{core}(M_1)$ in case of III(e,m,g) or CE) and $\tau_i$ is
the appropriate time scale.

%===================================================================
The radius of the star at stage III is assumed to follow the
RL radius:

\begin{equation}
\label{rd} R_l^d(M_d (t))=R_d (M_d (t)).
\end{equation}

\subsection{Wolf-Rayet and helium stars}

In the process of mass exchange the hydrogen envelope of the star
can be lost almost completely, so a hot white dwarf (for $m\le
2.5$), or a non-degenerate helium star (for higher masses)
% (a
% WR-star in case of a star with the initial mass $>10 M_{\odot}$)
is left as a remnant. The life-time of the helium star is determined by the
helium burning in the stellar core \citep{iben1985a}

\begin{equation}
\label{the} t_{He}=\left\{
\begin{array}{l}
1658 m^{-2}, \hspace{20pt} (\alpha) \\
1233 m^{-3.8}, \hspace{13pt} (\beta) \\
0.1 t_{H}, \hspace{35pt} (\gamma) \\
6913 m^{-3.47}, \hspace{10pt} (\delta) \\
\simeq 10, \hspace{38pt} (\epsilon) \\
0.1 t_H, \hspace{35pt} (\zeta)
\end{array}
\right.
\end{equation}

\noindent We assume the next indication in this formula:
($\alpha$), $m<1.1$, modes [IIA-IIF]; ($\beta$), $m>1.1$,
$m_{max}<10$, modes [IIA-IIF]; ($\gamma$), $m_{max}>10$, modes
[IIA-IIF]; ($\delta$), mode [IIIA]; ($\epsilon$), mode [IIIC];
($\zeta$), modes [IIIB,D,E] and type D.

If the helium (WR) star fills its Roche lobe (a relatively rare
so-called ``BB'' case of evolution; \citet{delgado1981a}; see
discussion in \citet{heuvel1994a}), the envelope is lost and a CO
stellar core is left with mass

\begin{equation}
\label{mc} m_c=\left\{
\begin{array}{l}
1.3+0.65(m-2.4), m\ge 2.5, \\
0.83 m^{0.36}, m<2.5, \\
\end{array}
\right.
\end{equation}

The mass-radius dependence in this case is \citep{tutukov1973a}

\begin{equation}
\label{mctut} r_{WR}=0.2m^{0.6}. \\
\end{equation}

\subsection{Stellar winds from normal stars}

The effect of the normal star on the compact magnetized component
is largely determined by the rate $\dot M$ and the velocity of stellar
wind at infinity $v_{\infty}$, which is assumed to be

\begin{equation}
\label{vinf} v_{\infty}=3v_p\approx 1.85\sqrt{m/r},
\end{equation}

\noindent where $v_p$ is the escape velocity at the stellar
surface.

For ``Be-stars'' (i.e. those stars at the stage ``I'' that
increased its mass during the first mass exchange), the wind
velocity at the infinity is taken to be equal to the Keplerian
velocity at the stellar surface:

\begin{equation}
\label{vbe} v_{\infty}=\sqrt{GM/R}\approx 0.44\sqrt{m/r}\,.
\end{equation}

\noindent The lower stellar wind velocity leads to an
effective increase of the captured mass rate by the secondary
companion to such ``Be-stars''.

The stellar wind mass loss rate at the stage ``I'' is calculated as

\citep{jager1980a}
\begin{equation}
\label{dm} \dot m=52.3\alpha_w l/v_{\infty},
\end{equation}

\noindent Here $\alpha_w=0.1$ is a numerical coefficient (in
general, we can treat it as free parameter).

For giant post-MS stars (stage ``II'') we assume $v_{\infty}=3v_p$
and for massive star we take the maximum between the stellar wind rate given
by de Jager's formula and that given by \citet{lamers1981a}

\begin{equation}
\label{lam} \dot m=\mbox{max}(52.3\alpha_w \frac{l}{v_{\infty}},
10^{2.33} \frac{l^{1.42}r^{0.61}}{m^{0.99}}),
\end{equation}
$$
M\ge 10 M_{\odot};
$$

For red super-giants we use Reimers's formula
\citep{kudritzki1978a}:

\begin{equation}
\label{kudr} \dot m=\mbox{max} (52.3 \alpha_w
\frac{l}{v_{\infty}}, 1.0 \frac{lr}{m}),
\end{equation}
$$
M\ge 10 M_{\odot};
$$

For a Wolf-Rayet star the stellar wind loss rate
can significantly increase (up to $10^{-5} M_{\odot}$
year$^{-1}$). We parametrize it as

\begin{equation}
\label{wrw} \dot M_{WR}=k_{WR}M_{WR}/t_{He},
\end{equation}

\noindent where the numerical coefficient is taken to
be $k_{WR}=0.3$ (in general, it can be changed if necessary).
The mass loss in other stages (MS, (super)giant) is assumed to be
limited  by 10\% of the mass of the star at the beginning of
the stage.

\subsection{Change of binary parameters: mass, semi-major axis and eccentricity}

The duration of any evolutionary stage is determined by the more rapidly evolving component
$\Delta t=\mbox{min}(\Delta t_1, \Delta t_2)$. On the other hand,
based on the evolutionary considerations we are able to calculate
how the mass of the faster evolving star changes
(e.g. due to the stellar wind or RL overflow), that is we can estimate the
quantity $\Delta M=M_i-M_f$. Then we set the
characteristic mass loss rate at this stage as

\begin{equation}
\label{dmo} \dot M_o=\Delta M/\Delta t,
\end{equation}

\noindent Next, we should calculate the change of mass for the
slower evolving companion, the orbital semi-major axis and the
eccentricity.

\subsection{Mass change}

The mass of the star loosing matter is calculated as

\begin{equation}
\label{mf} M_f=M_i-\dot M_o\times\Delta t,
\end{equation}
\noindent Accordingly, the mass of the accreting star is
\begin{equation}
\label{mf2} M_f=M_i+\dot M_c\times\Delta t,
\end{equation}

\noindent where $\dot M_c$ is the accretion rate of the captured
matter.

For stages without RL overflow the accretion rate of the captured
stellar wind matter is
\begin{equation}
\label{dmc} \dot m_c=3.8\times 10^{-2}
\left(\frac{m}{a(v_w^2+0.19(m_1+m_2)/a)}\right)^2 \dot m_o\,.
\end{equation}

At stages where RL overflow ocurs and both components
are normal (non-degenerate), we will assume that
the accretor can accomodate mass at the rate determined by
its thermal time, i.e.

\begin{equation}
\label{dmc2} \dot m_c=\dot m_o
\left(\frac{t_{KH}(donor)}{t_{KH}(accretor)}\right)\,.
\end{equation}

This means that the evolution can not be fully conservative, especially during
the first mass transfer where the primary
component usually has a shorter thermal time scale.

The mass
increase rate by compact accretors is assumed to be limited
by the critical Eddington luminosity (see, however, the possible hyper accretion
stage discussed below):

\begin{equation}
\label{edd} L_{Edd}=\frac{4\pi GM m_p}{\sigma_T}\approx1.3\cdot
10^{38}\times m \quad \mbox{erg/s}
\end{equation}

\noindent ($\sigma_T$ is the Thomson cross-section) {\it at
the stopping radius} $R_{stop}$ for the accreted matter (see, e.g.,
detailed discussion in \citet{lipunov1992a}). This corresponds to
the critical accretion rate

\begin{equation}
\label{crit} \dot M_{cr}=R_{stop} \frac{L_{Edd}}{GM}\,.
\end{equation}
\noindent Thus, the mass of the accreting compact star at the end
of the stage is determined by the relation
\begin{equation}
\label{dmf2} M_f=M_i+\mbox{min}(\dot M_c, \dot M_{cr})\times \Delta t\,.
\end{equation}

\subsection{Semi-major axis change}

The binary separation $a$ changes differently for various
mass exchange modes. First, we introduce a measure of
non-conservativeness of the mass exchange as the ratio between the
mass change of the accretor and the donor:
\begin{equation}
\label{beta}
\beta\equiv-(M_{accr}^i-M_{accr}^f)/(M_{donor}^i-M_{donor}^f).
\end{equation}

If the mass exchange is conservative ($\beta=1$, i.e.
$M_a+M_d=\mbox{const}$) and one can neglect
the angular momenta of the components, the orbital
momentum conservation implies
\begin{equation}
\label{s} \frac{a_f}{a_i}=\left(\frac{M_a^i M_d^i}{M_a^f
M_d^f}\right)^2\,.
\end{equation}

In a more general case of quasi-conservative mass transfer
$0\le\beta<1$, the orbital separation changes differently
depending on the specific angular momentum carried away from the
system by the escaping matter (see
\citet{heuvel1994a} for more detail).
We treat the quasi-conservative mass transfer by assuming the isotropic mass
loss mode in which the matter carries away the specific orbital
angular momentum of the accreting component ($j_a$)

\begin{equation}
\label{jout} \dot J_{out}=(1-\beta)\dot M_c j_a\,.
\end{equation}
\noindent From here we straightforwardly find
\begin{equation}
\label{semiaxis}
\frac{a_f}{a_i}=\left(\frac{q_f}{q_i}\right)^3\left(\frac{1+q_i}{1+q_f}\right)\left(\frac{1+\frac{\beta}{q_f}}{1+\frac{\beta}{q_i}}\right)^{3+2/\beta}\,.
\end{equation}

\noindent In this formula $q=M_{accr}/M_{donor}$ and
the non-conservative
parameter $\beta$ is set to be the minimal value between $\beta=1$ and the
ratio $T_{KH}(donor)/T_{KH}(accr)$ ($T_{KH}(donor)$ and $T_{KH}(accr)$ is
the thermal time of donor and accretor, respectively).

When no matter is captured by the secondary companion without
additional losses of angular momentum (the so-called ``absolutely
non-conservative case''), which relates to the spherical-symmetric
stellar wind from one component, we use another well-known formula

\begin{equation}
\label{nc} \frac{a_f}{a_i}=\frac{M_1^i+M_2^i}{M_1^f+M_2^f}\,.
\end{equation}
\noindent In this case the orbital separation always increases.

When the orbital angular momentum is carried away by GW or MSW
with no RL overflow, the following approximate formulas are used:

\begin{equation}
\label{mswgw} \frac{a_f}{a_i}=\left\{
\begin{array}{l}
(1-\Delta t/\tau_{MSW})^{1/4}, \mbox{for MSW}, \\
(1-\Delta t/\tau_{GW})^{1/5}, \mbox{for GW},
\end{array}\right.
\end{equation}

In a special case of a white dwarf filling its RL (stage ``IIIwd''
above), assuming a stable (i.e. where $d \ln
R_{wd} / d \ln M= d \ln R_{RL}/ d \ln M$) conservative mass
transfer with account for the mass-radius relation $R_{wd}\propto
M_{wd}^{-1/3}$, the orbital separation must
increase according to the equation

\begin{equation}
\label{cons} \frac{a_f}{a_i}=\left(\frac{m_f}{m_i}\right)^{-2/3},
\end{equation}

\noindent where $m_i$ is the initial mass of the WD donor and $m_f$ is its
mass at the end of the mass transfer.

\subsection{The change of eccentricity}

Tidal interaction between components, as well as the orbital
angular momentum loss due to MSW or GR decrease the eccentricity
of the binary system. The tidal interaction is essential in very
close binaries or even during the common envelope stage. MSW is
effective only in systems with low-mass late-type main sequence
stars (see above), GW losses become significant only in
short-period binaries.

The tidal interaction conserves the orbital angular momentum which
implies the relation

\begin{equation}
\label{eccen} a(1-e^2)=\mbox{const}.
\end{equation}

%======================================================
It seems that the orbit becomes a circle faster than major
semi-axis of the orbit decreases during common envelope stage. We
suppose that $t_{cyr}=1/3 t_{CE}$. We accept that $t_{cyr}$ for
RL-filling stars is equal to its Kelvin-Helmholtz time. Detached
systems change their eccentricity during the next character time
\citep{zahn1975a, press1977a, zahn1989a, zahn1989b}

\begin{equation}
\label{tc}
t_{cyr}=t_{KH}\left(\frac{1+e}{1-e}\right)^{3/2}\left(\frac{R}{R_L}\right)^{-5}.
\end{equation}

\noindent here $t_{KH}\approx GM^2/R_L$ is thermal time of the
star, $R$ is radius of the star, $R_L$ is its RL size. For systems
which consist of two normal stars we choose minimal value of
$t_{cyr}$. Resonances at very high eccentricities are not taken
into account \citep{mardling1995a, mardling1995b}.

Orbits of the systems with MSW become circular during
$t_{cyr}=\tau_{MSW}$ (see \ref{mswgw}).

In case of GW analytical exact solutions were obtained for $a(t)$
and $e(t)$ \citep{peters1963a, peters1964a}.

%=============================================
\section{Special cases: supernova explosion and common envelope}

\label{kick}

Supernova explosion in a binary is treated as an instantaneous
mass loss of the exploding star. The additional kick
velocity can be imparted to the newborn neutron star due
the collapse asymmetry (see below for discussion). In this case
the eccentricity and semi-major axis of the binary after the
explosion can be straightforwardly calculated
\citep{boersma1961a} (see necessary formulas also in Grishchuk et al. 2001).
Briefly, we use the following scheme.

\begin{enumerate}

\item First, velocities and locations of the components on the
orbit prior to the explosion are calculated;

\item then the mass of the exploding star $M_{pr}$ - $M_{remnant}$
is changed and the arbitrarily directed
kick velocity $w$ is added to its orbital velocity;

\item after that the transition to the new system's barycenter is performed
(at this point the spatial velocity of the new center of mass
of the binary is calculated);

\item in this new reference frame the new total energy
$E'_{tot}$ and the orbital angular momentum $J'_{orb}$ are computed;
if the new total energy is negative, the new semi-major axis
$a'$ and eccentricity $e'$ are calculated by using the new
$J'_{orb}$ end $E'_{tot}$; if the total energy is positive (that
is, the binary is unbound) spatial velocities of each component
are calculated.
\end{enumerate}

% It was supposed in our calculations that a neutron star or a black
% hole during supernova explosion can get additional ``kick'', its
The kick velocity $w$ distribution is taken in the Maxwellian form:
\begin{equation}
f(w)\sim \frac{w^2}{w^2_0}e^{-\frac{w^2}{w_0^2}}\,. \label{Maxwell}
\end{equation}

We suppose that the absolute value of the velocity that can be
added during the formation of a black hole depends on the mass
loss by the collapsing star, the value of the parameter $w_0$
during the BH formation is defined as

\begin{equation}
\label{w0bh} w_{0}^{bh}=\left(1-k_{bh}\right)w_0.
\end{equation}

An effective spiral-in of the binary components occurs during the
common envelope (CE) stage. This complicated process (introduced
by \citet{paczynski1976a}) is not fully understood as yet, so we
use the conventional energy consideration to find the binary
system parameters after the CE by introducing a parameter
$\alpha_{CE}=\Delta E_b/\Delta E_{orb}$, where $\Delta
E_b=E_{grav}-E_{thermal}$ is the binding energy of the ejected
envelope matter and $\Delta E_{orb}$ is the drop in the orbital
energy of the system during the spiral-in phase \citep{heuvel1994a}. This
parameter measures the fraction of the system's orbital energy
% (between the beginning and the end of the spiral-in process)
that comes during the spiral-in process to the binding energy
(gravitational minus thermal) of the ejected common envelope. Thus

\begin{equation}
\alpha_{CE} \left( \frac{GM_a M_c}{2a_f} - \frac{GM_a M_d}{2a_i}
\right)=\frac{GM_d(M_d - M_c)}{R_d},
\end{equation}

\noindent where $M_c$ is the mass of the core of the mass-losing
star with the initial mass $M_d$ and radius $R_d$ (which is simply
a function of the initial separation $a_i$ and the initial mass
ratio $M_a/M_d$, where $M_a$ is the mass of the accreting star).

On the CE stage the luminosity of the accreting star can reach the
Eddington limit so that the further increase of the accretion rate
can be prevented by radiation pressure. This usually happens at
accretion rates $\dot M \simeq 10^{-4} - 10^{-5} M_{\odot}$
yr$^{-1}$. However, \citet{chevalier1993a} suggested that when the
accretion rate is higher ($\dot M \simeq 10^{-2} - 10^{-3}
M_{\odot}$ yr$^{-1}$), the energy is radiated away not by
high-energy photons only, but also by neutrinos (see also
\citet{zeldovich1972a} and the next section). On the typical time
scale for the hyper accretion stage of $10^2$ yr, up to $\sim
1M_{\odot}$ of matter may be incident onto the surface of the
neutron star.

\section{Three regimes of mass accretion by neutron stars}

A considerable fraction of observed neutron stars have increased
their masses in the course of their evolution, or are still
increasing their masses (e.g., in X-ray sources). But how large
can this mass increase be? It is clear that the only origin of a
mass increase is accretion. It is evident that the overall change
in the mass of a neutron star is determined not only by the
accretion rate, but also by the duration of the accretion stage:

\begin{equation}
\label{dm} \Delta M=\int_{0}^{T_a} \dot M dt=\dot M T_a,
\end{equation}

\noindent where $\dot M$ is the mean accretion rate and $T_a$ is
the lifetime of the accretion stage. We emphasize that, in the
case under consideration, the accretion rate is the amount of
matter falling onto the surface of the neutron star per unit time,
and can differ signifcantly from the values indicated by the
classical Bondi-Hoyle formulas. Three regimes of accretion are
possible in a close binary containing a neutron star: ordinary
accretion, super-accretion, and hyper-accretion.

\subsection{Ordinary accretion}

The ordinary accretion regime is realized when all matter captured
by the gravitational field of the neutron star falls onto its
surface. This is possible only if the radiation pressure and
electromagnetic forces associated with the magnetic field of the
star and its rotation are small compared to the gravitational
force. In this case, the increase in the mass will be precisely
determined by the gas dynamics of the accretion at the
gravitational-capture radius or, if the donor fills its Roche
lobe, by the binary mass ratio and the evolutionary status of the
optical component. In this case, the accretor is observed as an
X-ray source with luminosity

\begin{equation}
\label{rllum} L_x=\dot M \frac{GM_x}{R_*},
\end{equation}

\noindent where $M_x$ and $R_*$ are the mass and radius of the
neutron star. The accretion rate $\dot M$ is determined by the
Bondi-Hoyle formula

\begin{equation}
\label{bondihoyle} \dot M= \pi R_G^2\rho v\,,
\end{equation}

\noindent where $R_G$ is the gravitational-capture radius of the
neutron star, $v$ is the velocity of the gas flow relative to the
neutron star, and $\rho$ is the density of the gas.

The X-ray luminosity of the accretor $L_x$ and its other main
parameters can be used to estimate the mass $\Delta M$ accumulated
during the accretion phase:
\begin{equation}
\label{deltam} \Delta M=\frac{L_x R_* T_a}{G M_x},
\end{equation}

\subsection{Super-accretion}

Regime of super accretion was considered, for instance, in paper
\citet{lipunov1982d}. Despite the absence of detailed models for
supercritical disk accretion (supercritical accretion is realized
precisely via an accretion disk), it is possible to estimate the
main characteristics of the process -- the accretion rate,
magnetosphere radius, and evolution equations. Accretion is
considered to be supercritical when the energy released at the
radius where the accretion exceeds the Eddington limit:

\begin{equation}
\label{sa} \dot M\frac{G M_x}{R_{stop}}>L_{Edd}=1.38\times
10^{38}(M_x/M_{\odot}) \quad \mbox{erg s$^{-1}$},
\end{equation}

\noindent where $R_{stop}$ is either the radius of the neutron
star or the magnetosphere radius $R_A$.

For strongly magnetized neutron stars with magnetic fields $B\gg
10^8$ G, all matter arriving at the magnetosphere is accreted onto
the magnetic poles, where the corresponding gravitational energy
is released. If the black body temperature $T$, roughly estimated
as

\begin{equation}
\label{bb} S\sigma T^4=\dot M \frac{G M_x}{R_*},
\end{equation}

\noindent is higher than $5\times 10^9$ K ($S$ is the area of the
base of the accretion column), most of the energy will escape from
the neutron star in the form of neutrinos, and, hence, will not
hinder accretion \citep{zeldovich1972a,basko1975a}. In this case,
the rate at which the neutron star accumulates mass will be

\begin{equation}
\label{rt} \dot M\simeq \dot M_{crit}
\left(\frac{R_A}{R_*}\right)^2\gg \dot M_{crit},
\end{equation}

For lower temperatures there should be an upper limit on
the accretion rate equal to the standard Eddington limit.

\subsection{Hyper-accretion}

A considerable fraction of neutron stars in binary systems pass
through the common-envelope stage in the course of their
evolution. In this case, the neutron star is effectively immersed
in its optical companion, and for a short time ($10^2$-$10^4$ yrs)
spirals-in inside a dense envelope of the companion. The formal
accretion rate estimated using the Bondi-Hoyle formulas is four to
six orders of magnitude higher than the critical rate and, as was
suggested by \citet{chevalier1993a}, this may result in
hyper-accretion, when all the energy is carried away by neutrinos
for the reasons described above. There are currently no detailed
theories for hyper-accretion or the common-envelope stage. The
amount of matter accreted by the neutron star can be estimated as

\begin{equation}
\label{hyp} \Delta
M=\int^{T_{hyper}}_{0}\frac{1}{4}\left(\frac{R_G}{a}\right)^2\dot
M dt\simeq
\end{equation}
$$
\simeq\frac{1}{4}(M_{opt}-M_{core})\left(\frac{M_x}{M_{opt}}\right)^2;
$$

\noindent where $T_{hyper}$ is the duration of the hyper-accretion
stage, $R_G$ is the gravitational-capture radius of the neutron
star, $a$ is the initial semi-major axis of the close binary
orbit, $M_{core}$ is the mass of the core of the optical star, and
$M_{opt}$ and $M_x$ are the total masses of the optical star and
of the neutron star at the onset of the hyper-accretion stage. The
mass of the neutron star can increase during the common envelope
stage as much as $\sim 1 M_{\odot}$ \citep{bogomazov2005a}, up to
$M_{OV}$. Such NSs collapse into black holes.

\section{Mass accretion by black holes}

If the black hole has formed in the binary system, its X-ray
luminosity is

\begin{equation}
\label{bhlum} L_x=\mu \dot M c^2,
\end{equation}

\noindent where $\mu=0.06$ for Schwarzschild black hole and
$\mu=0.42$ (maximum) for extremal Kerr BH.

We use the Bondi-Hoyle formulas to estimate the accretion rate
onto BH.

Powerful X-ray radiation is able to originate only if an accretion
disc has formed around the black hole \citep{karpov}. For
spherically symmetric accretion onto a black hole the X-ray
luminosity is insignificant. A very low stellar wind velocity is
necessary to form an accretion disc \citep{lipunov1992a}

\begin{equation}
\label{windvel} V<V_{cr}\approx
\end{equation}
$$
\approx 320 (4\eta)^{1/4} m^{3/8} T_{10}^{-1/4} R_8^{-1/8}
(1+\tan^2 \beta)^{-1/2},
$$

\noindent where $\eta$ is averaged over the z-coordinate dynamic
viscous coefficient, $m=M_x/M_{\odot}$, $M_x$ is the relativistic
star mass, $T_{10}=T/10$, $T$ is the orbital period in days,
$R_8=R_{min}/10^8\mbox{cm}$, $R_{min}$ is the minimal distance
from the compact object up to which free Keplerian motion is still
possible and $\beta$ is the accretion disk axis inclination angle
with respect to the radial direction. For black holes
$R_{min}=3R_g$, where $R_g=2GM_{bh}/c^2$.

\section{Accretion induced collapse and compact objects merging}

WD and NS are degenerate configurations, which have upper limit of
their mass (the Chandrasekhar and Oppenheimer-Volkov limits
correspondingly). The Chandrasekhar limit depends on chemical
composition of the white dwarf
\begin{equation}
\label{mchl} M_{Ch}=\left\{\begin{array}{l}
1.44 M_{\odot}, \quad \mbox{He WD}, \\
1.40 M_{\odot}, \quad \mbox{CO WD}, \\
1.38 M_{\odot}, \quad \mbox{ONeMg WD},
\end{array}\right.
\end{equation}

If the mass of WD becomes equal to $M_{Ch}$, the WD loses
stability and collapses. The collapse is accompanied by the
powerful thermonuclear burst observed as a type Ia supernova.
Collapses of He and CO WDs leave no remnants \citep{nomoto1991a}.
The outcome of the collapse of a ONeMg WD is not clear. It can
lead to the formation of a neutron star (the accretion induced
collapse, AIC). Some papers come to the different conclusion about
the result of AICs (see e.g. \citet{gb1994,ritossa1996a}). The
question about the NS formation during the WD collapse remains
undecided. Nevertheless, some NSs could have been formed from AIC
WD \citep{paradijs1997a}. In the ``Scenario Machine'' code the
possibility of NS formation during ONeMg WD is optional.

The merging of two compact objects in a binary WD system (e.g.,
due to the GW losses) should likely to be similar to AIC. During
the merging of two helium WDs,  one object with a mass of less
than $M_{Ch}$ (for example, $0.5M_{\odot}+0.6M_{\odot}\to
1.1M_{\odot}$) can form. At the same time, if the total mass of
the components exceeds $1.0-1.2 M_{\odot}$, a thermonuclear
burning can happen. It is likely that the merging of a ONeMg WD
with another WD can form a NS.

For the typical NS mass $\approx 1.4M_{\odot}$, the binary NS+NS
merging event can produce a black hole. Massive ($\approx
2.8M_{\odot}$) neutron star can be formed only if the NS equation
of state is very hard and $M_{OV}\simeq 2.8-3.0 M_{\odot}$.

BH+BH merging should produce a rapidly rotating black hole with
the mass equal to the total mass of the coalescing binary.

\section{Additional scenarios of stellar wind from massive stars}

\subsection{Evolutionary scenario B}

In the end of 1980s and in the beginning of 1990s the series of
new evolutionary tracks were calculated. The authors used new
tables of opacities \citep{rogers1991a,kurucz1991a}, new
cross-sections in nuclear reactions \citep{landre1990a} and new
parameters of convection in stars \citep{stothers1991a}. For stars
with $M<10M_{\odot}$ those tracks proved to be almost coincident
with previous calculations. More massive stars had much stronger
stellar winds.

A massive star loses up to 90\% of its initial mass in the
main-sequence, supergiant, and Wolf-Rayet stages via stellar wind.
Therefore, the presupernova mass in this case can be $\approx
8-10M_{\odot}$, essentially independent of the initial mass of the
star \citep{jager1988a,jager1990a,schaller1992a}.

\subsection{Evolutionary scenario C}

The papers mentioned above were criticised and in 1998 a new
version of the evolutionary scenario was developed
\citep{vanbeveren1998a}. The stellar wind loss rates were
corrected taking into account empirical data about OB and WR
stars. Here we list the main equations of this scenario (all
results concern only with the stars with initial mass $M_0>15
M_{\odot}$).

% \textbf{Significant differences are in stellar wind and boundary
% condition for stellar evolution:}

\begin{equation}
\label{wrw} \log \dot M=\left\{
\begin{array}{l}
1.67\log L-1.55\log T_{eff}-8.29, \quad(\alpha) \\
\log L+\log R-\log M-7.5,\quad(\beta)
\\
0.8\log L -8.7, \quad(\gamma)\\
\log L-10, \quad(\delta)
\end{array}\right.
\end{equation}

\noindent We assume the next indication in this formula:
($\alpha$), H burning in the core; ($\beta$), giant, $M_0\ge
40M_{\odot}$; ($\gamma$), giant, $M_0< 40M_{\odot}$; ($\delta$),
Wolf-Rayet star.

Note that in this subsection we used the mass $M$ is in
$M_{\odot}$, the luminosity $L$ is in $L_{\odot}$ \ and the radius
$R$ is in $R_{\odot}$. With these new calculations, the Webbink
diagram described above changed significantly
\citet{vanbeveren1998a}.

In this scenario, the total mass loss by a star is calculated
using the formula

\begin{equation}
\Delta M=(M-M_{core}), \label{m_A}
\end{equation}

\noindent where $M_{core}$ is the stellar core mass
(\ref{m_core_C}). If the maximum mass of the star (usually it is
the initial mass of a star, but the mass transfer in binary
systems is able to increase its mass above its initial value)
$M_{max}>15M_{\odot}$, the mass of the core in the main sequence
stage is determined using (\ref{m_core_C}$\alpha$), and in giant
and supergiant stages using (\ref{m_core_C}$\beta$). In the
Wolf-Rayet star stage (helium star), if $M_{WR}<2.5M_{\odot}$ and
$M_{max}\le 20M_{\odot}$ it is described using
(\ref{m_core_C}$\gamma$), if $M_{WR}\ge 2.5M_{\odot}$ and
$M_{max}\le 20M_{\odot}$ as (\ref{m_core_C}$\delta $), if
$M_{max}> 20M_{\odot}$ as (\ref{m_core_C}$\epsilon$)

\begin{equation}
m_{core} = \left\{
\begin{array}{l}
1.62m_{opt}^{0.83} \quad\quad\quad\quad\quad\quad\quad\quad  \text{($\alpha$)} \\
10^{-3.051+4.21\lg m_{opt} -0.93(\lg m_{opt})^2}
\quad \text{($\beta$)} \\
0.83m_{WR}^{0.36} \quad\quad\quad\quad\quad\quad \text{($\gamma$)} \\
1.3+0.65(m_{WR}-2.4)
\quad \text{($\delta$)} \\
m_{core}=3.03m_{opt}^{0.342} \text{($\epsilon$)} \label{m_core_C}
\end{array}\right.
\end{equation}

These evolutionary scenarios have some peculiar properties with
respect to the classical scenario. One of them is that the strong
stellar wind from massive stars leads to a rapid and significant
increase of the system's orbit and such stars cannot fill its RL
at all.

There are three observational facts that conflict with the strong
stellar wind scenarios:

\begin{enumerate}

\item A very high $\dot M$ is a problem by itself. The observers
calculate this quantity for most of OB and WR stars using the
emission measure $EM\propto \int n_e^2 dl$. The estimate of $\dot
M$ using EM is maximal for homogenous wind. However, there are
evidences (see e.g. \citet{cherepashchuk1984a}) that stellar winds
of massive stars are strongly ``clumpy''. In this case the real
$\dot M$ must be 3-5 times less. This note is especially important
for the scenario with high stellar wind.

\item Very massive Wolf-Rayet stars do exist. There are at least
three double WR+OB systems including very massive WR-stars: CQ Cep
$40 M_{\odot}$, HD 311884 $48 M_{\odot}$ and HD92740 $77
M_{\odot}$ \citep{cherepashchuk1996a}. Such heavy WR stars are at
odds with the assumed high mass loss rate.

\item In the semi-detached binary system RY Sct (W Ser type) the
mass of the primary component is $\approx 35 M_{\odot}$
\citep{cherepashchuk1996a}. This mass is near the limit \\
\citep{vanbeveren1998a} beyond which the star, according to the
high mass loss scenario, cannot fill its RL.

\end{enumerate}

\subsection{Evolutionary Scenario W}

The evolutionary scenario W is based on the stellar evolution
calculations by Woosley et al. (\citet{woosley2002a}, Fig. 16),
which represents the relationship between the mass of the
relativistic remnant and the initial mass of the star. We included
into population-synthesis code two models with W-type stellar
winds, which we label Wb and Wc. In models Wb and Wc, the
mass-loss rates were computed as in scenario B and scenario C,
respectively. The use of these models to calculate the wind rate
in a scenario based on Woosley's diagram (\citet{woosley2002a},
Fig. 16) is justified by the fact that scenarios B and C are based
on the same numerical expressions for the mass-loss rates from
\citet{schaller1992a,vanbeveren1998a,jager1990a} that were used by
\citet{woosley2002a}.

\section{The ``Ecology'' of Magnetic Rotators}

One of the most important achievements in astrophysics in
the end of the 1960s was the realization that in addition to
``ordinary'' stars, which draw energy from nuclear
reactions, there are objects in the Universe whose radiation is
caused by a strong gravitational and magnetic field. The well-known
examples include neutron stars and white dwarfs. The property that
these objects have in common is that their astrophysical
manifestations are primarily determined by interaction with the surrounding
matter.

In the early 1980s, this approach led to the creation of a
complete classification  scheme involving various regimes of
interaction between neutron stars and their environment, as well
as to the first Monte Carlo simulation of the NS evolution
\citep{lipunov1984a}. In addition to NSs, this scheme has been shown to
be applicable to other types of magnetized rotating stars.

By virtue of the relationship between the gravitational and
electromagnetic forces, the NS in various states can manifest
itself quite differently from the astronomical point of view.
Accordingly, this leads to the corresponding classification of NS
types and to the idea of NS evolution as a gradual changing of
regimes of interaction with the environment. The nature of the NS
itself turns out to be important also when constructing the
classification scheme. This indicates that there should be a whole
class of quite different objects which have an identical physical
nature. To develop the theory describing properties of such
objects (in a sense, it should establish ``ecological'' links
between different objects), it proved to be convenient to use symbolic
notations elaborated for the particular case of NS. We start this
subsection with recollecting the magnetic rotator formalism
(mainly according to the paper \citet{lipunov1987a}).

\subsection{A Gravimagnetic Rotator}

We call any gravitationally-bounded object having an angular
momentum and intrinsic magnetic field  by the term ``gravimagnetic
rotator''  or simply, rotator. In order to specify the intrinsic
properties of the rotator, three parameters are sufficient -- the
mass $M$, the total angular momentum $J=I\omega$ ($I$ is the
moment of inertia and $\omega$ is the angular velocity), and the
magnetic dipole moment $\mu$. Given the rotator radius $R_0$, one
can express the magnetic field  strength at the poles $B_0$ by
using the dipole moment $B_0=2\mu/R_0^3$. The angle $\beta$
between the angular moment $\bf J$ and the magnetic dipole moment
$\mu$ can also be of importance: $\beta=\arccos(J\mu)$.

\subsection{The Environment of the Rotator}

We assume that the rotator is surrounded by an ideally conductive
plasma with a density $\rho_{\infty}$ and a sound velocity
$a_{\infty}$ at a sufficiently far distance from the rotator. The
rotator moves relative to the environment with a velocity
$v_{\infty}$. Under the action of gravitational attraction, the
surrounding matter should fall onto the rotator. A rotator
without a magnetic field would capture a stationary flow of
matter, $\dot M_c$, which can be estimated using the Bondy-Hole-Lyttleton
formulae \citep{bondi1944a,bondi1952a,mccrea1953a}:

\begin{equation}
\label{bhl} \dot
M_c=\delta\frac{(2GM)^2}{(a^2_{\infty}+v^2_{\infty})^{3/2}}\rho_{\infty},
\end{equation}

\noindent where $\delta$ is a dimensionless factor of the order of
unity. When one of the velocities, $a_{\infty}$ or $v_{\infty}$,
far exceeds the other, the accretion rate  is determined by the
dominating velocity, and can be written in a convenient form as
(\ref{bondihoyle}).

In the real astrophysical situation, the parameters of the
surrounding matter at distances $R\gg R_G$ can be taken as
conditions at infinity\footnote{$R_G=\frac{2GM}{v^2}$.}.

As already noted, the matter surrounding a NS or a WD is almost
always in the form of a high-temperature plasma with a high
conductivity. Such accreting plasma must interact efficiently with
the magnetic field of the compact star  \citep{amnuel1968a}.
Hence, the interaction between the compact star and its
surroundings cannot be treated as purely gravitational and
therefore the accretion  is not a purely gas dynamic process. In
general, such interaction should be described by the magneto
hydrodynamical equations. This makes the already complicated
picture of interaction of the compact star with the surrounding
medium even more complex.

The following classification  of magnetic rotators   is based on
the essential characteristics of the interaction of the plasma
surrounding them with their electromagnetic field. This approach
was proposed by \citet{shvartsman1970a} who distinguished three
stages of interaction of magnetic rotators: the ejection stage,
the propeller stage,  which was later rediscovered by
\citet{illarionov1975a} and named as such, and the accretion
stage. Using this approach, \citet{shvartsman1971a} was able to
predict the phenomenon of accreting X-ray pulsars  in binary
systems. New interaction regimes discovered later have led to a
general classification of magnetic rotators
\citep{lipunov1982a,lipunov1984a,kornilov1983a}.

It should be noted that the interaction of the magnetic rotator
with the surrounding plasma is not yet understood in detail.
However, even the first approximation reveals a multitude of
interaction models. To simplify the analysis, we assume the
electromagnetic part of the interaction to be independent of the
accreting flux parameters, and vice versa.

Henceforth, we shall assume in almost all cases that the intrinsic
magnetic field of a rotator is a dipole field \citep{landau1971a}:

\begin{equation}
\label{df} B_d=\frac{\mu}{R^3}(1+3\sin^2 \theta)^{1/2},
\end{equation}

This is not just a convenient mathematical simplification. We will
show that the magnetoplasma interaction takes place at large
distances from the surface of the magnetic rotator,  where the
dipole moment makes the main contribution. Moreover, the collapse
of a star into a NS is known to ``cleanse'' the magnetic field.
Indeed, the conservation of magnetic flux leads to a decrease of
the ratio of the quadrupole magnetic moment $q$ to the dipole
moment $\mu$ in direct proportion to the radius of the collapsing
star, $q/\mu\propto R$.

It should be emphasized, however, that the contribution of the
quadrupole component to the field strength at the surface remains
unchanged.

The light cylinder radius  is the first important characteristric
of the rotating magnetic field:
\begin{equation}
\label{lc} R_l=\frac{c}{\omega},
\end{equation}
\noindent where $c$ is the speed of light.

A specific property of the field of the rotating magnetic dipole
in vacuum is the stationarity of the field inside the light
cylinder and formation of magneto dipole radiation beyond the
light cylinder. The luminosity of the magnetic dipole  radiation
is equal to \citep{landau1971a}:

\begin{equation}
\label{landausifshits}
L_m=\frac{2}{3}\frac{\mu^2\omega^4}{c^3}\sin^2
\beta=k_t\frac{\mu^2}{R_l^3}\omega,
\end{equation}

\noindent where $k_t=\frac{2}{3}\sin^2\beta$.

This emission exerts a corresponding braking torque

\begin{equation}
\label{ve} {\bf K_m}=- \left(\frac{2}{3}\right)\frac{\mu^2
\sin^2\beta}{c^3}\omega^3 {\bf c_{\omega}},
\end{equation}

leading to a spin down of the rotator. Although magnetic dipole
radiation from pulsars do not exists, almost all models predict
energy loss quantity near this value. At the same time we do not
take into account possibly complicated angular dependence of such
loss.

\subsection{The Stopping Radius}

Now we consider qualitatively the effect of the electromagnetic
field of a magnetic rotator on the accreting plasma. Consider a
magnetic rotator with a dipole magnetic moment $\mu$, , rotational
frequency $\omega$, and mass $M$. At distances $R\gg R_G$ the
surrounding plasma is characterized by the following parameters:
density $\rho_{\infty}$, sound velocity $a_{\infty}$ and/or
velocity $v_{\infty}$ relative to the star. The plasma will tend
to accrete on to the star under the action of gravitation. The
electromagnetic field, however, will obstruct this process, and
the accreting matter will come to a stop at a certain distance.

Basically, two different cases can be considered:

\begin{itemize}

\item When the interaction takes place beyond the light cylinder,
$R_{stop}>R_l$. This case first considered by
\citet{shvartsman1970a,shvartsman1971a}. In this case the magnetic
rotator generates a relativistic wind consisting of a flux of
different kinds of electromagnetic waves and relativistic
particles. The form in which the major part of the rotational
energy of the star is ejected is not important at this stage. What
is important is that both relativistic particles and magnetic
dipole radiation will transfer their momentum and hence exert
pressure on the accreting plasma. Indeed, random magnetic fields
are always present in the accreting plasma. The Larmor radius of a
particle with energy $\ll 10^{10}$ eV moving in the lowest
interstellar magnetic field $\sim 10^{-6}$ G is much smaller than
the characteristic values of radius of interaction, so the
relativistic wind will be trapped by the magnetic field of the
accreting plasma and thus will transfer its momentum to it.

Thus, a relativistic wind can effectively impede the accretion of
matter. A cavern is formed around the magnetic rotator,   and the
pressure of the ejected wind $P_m$ at its boundary balances the
ram pressure of the accreting plasma $P_a$:

\begin{equation}
\label{pmpa} P_m(R_{stop})=P_a(R_{stop}),
\end{equation}

This equality defines a characteristic size of the stopping
radius, which we call the Shvartzman radius $R_{Sh}$.

\item The accreting plasma penetrates the light cylinder
$R_{stop}<R_l$. The pressure of the accreting plasma is high
enough to permit the plasma to enter the light cylinder. Since the
magnetic field inside the light cylinder decreases as a dipole
field, the magnetic pressure is given by

\begin{equation}
\label{pm} P_m=\frac{B^2}{8\pi}\approx\frac{\mu^2}{8\pi R^6},
\end{equation}

Matching this pressure to the ram pressure of the accreting plasma
yields the Alfven radius $R_A$.

The magnetic pressure and the pressure of the relativistic wind
can be written in the following convenient form:

\begin{equation}
\label{pm2} P_m= \left\{\begin{array}{l}
\frac{\mu^2}{8\pi R^6}, \quad R\le R_l, \\
\frac{L_m}{4\pi R^2 c}, \quad R>R_l, \end{array}\right.
\end{equation}

We introduce a dimensionless factor $k_t$ such that the power of
the ejected wind is

\begin{equation}
\label{lm} L_m=k_t \frac{\mu^2}{R_l^3}\omega,
\end{equation}

Assuming $k_t=1/2$ we get for $R=R_l$ a continuous function
$P_m=P_m(R)$.

The accreting pressure of plasma outside the capture radius  is
nearly constant, and hence gravitation does not affect the medium
parameters significantly. In contrast, at distances inside the
gravitational capture radius $R_G$ the matter falls almost freely
and exerts pressure on the ``wall'' equal to the dynamical
pressure. For spherically symmetric accretion we obtain

\begin{equation}
\label{pa} P_a=\left\{\begin{array}{l}
\frac{\dot M v_{\infty}}{4\pi R_G^2}, \quad R>R_G, \\
\frac{\dot M v_{\infty}}{4\pi
R_G^2}\left(\frac{R_G}{R}\right)^{1/2}, \quad R\le R_G,
\end{array}\right.
\end{equation}

Here we used the continuity equation $\dot M_c=4\pi R_G^2
\rho_{\infty}v_{\infty}$. When presented in this form, the
pressure $P_a$ is a continuous function of distance.

Summarizing, for the stopping radius  we get

\begin{equation}
\label{pa} R_{stop}=\left\{\begin{array}{l}
R_a, \quad R_{stop}\le R_l, \\
R_{Sh}, \quad R_{stop}>R_l,
\end{array}\right.
\end{equation}

The expressions for the Alfven radius  are:

\begin{equation}
\label{pa} R_A=\left\{\begin{array}{l}
\left(\frac{2\mu^2 G^2 M^2}{\dot M_c v_{\infty}^5}\right)^{1/6}, \quad R_A>R_G, \\
\left(\frac{\mu^2}{2\dot M_c \sqrt{2GM}}\right)^{2/7}, \quad
R_A\le R_G,
\end{array}\right.
\end{equation}

and for the Shvartzman radius:

\begin{equation}
\label{rsh} R_{Sh}=\left( \frac{8k_t\mu^2G^2m^2\omega^4}{\dot M_c
v^5_{\infty} c^4} \right)^{1/2}, R_{Sh}>R_G,
\end{equation}

\end{itemize}

\subsection{The Stopping Radius in the Supercritical Case}

The estimates presented above for the stopping radius were
obtained under the assumption that the energy released during
accretion does not exceed the Eddington limit, so we neglected the
reverse action of radiation on the accretion flux parameters.

Now, we turn to the situation where one cannot neglect the
radiation pressure. Consider this effect after
\citet{lipunov1982b}. Suppose that the accretion  rate of matter
captured by the magnetic rotator is such that the luminosity at
the stopping radius  exceeds the Eddington limit (see equation
\ref{edd}).

We shall assume, following \citet{shakura1973a}, that the radiation
sweeps away exactly that amount of matter which is needed for the
accretion luminosity  of the remaining flux to be of the order of
the Eddington luminosity  at any radius:

\begin{equation}
\label{shsun} \dot M(R)\frac{GM}{R}=L_{edd},
\end{equation}

This yields
\begin{equation}
\label{gv} \dot M(R)=\dot M_c\frac{R}{R_s}, \quad
R_s=\frac{\kappa}{4\pi c}\dot M_c,
\end{equation}

\noindent where $R_s$ is a spherization radius (where the
accretion luminosity first reaches the Eddington limit), and
$\kappa$ designates the specific opacity of matter. Using the
continuity equation, the ram pressure of the accreting plasma is
now obtained as another function of the radial distance

\begin{equation}
\label{rmp} P_a\approx \rho v^2\approx \frac{\dot M(R)}{4\pi
R^2}v=\frac{\dot M_c\sqrt{2GM}}{4\pi R_s}R^{-3/2}, \quad R\le R_s,
\end{equation}

in contrast to the subcritical regime when $P_a\propto R^{-5/2}$.
Matching $P_a$ and $P_m$ (see the previous section) for the
supercritical case gives

\begin{equation}
\label{pmpas} R_A=\left(\frac{\mu^2\kappa}{8\pi
c\sqrt{2GM_x}}\right)^{2/9},
\end{equation}
\begin{equation}
R_{Sh}=\left(\frac{\kappa k_t\mu^2\omega^4}{4\pi c^5
\sqrt{2GM}}\right)^{2},
\end{equation}

The critical accretion rate $\dot M_{cr}$ is defined by the
boundary of the inequality
\begin{equation}
\label{mcr} \dot M_c\ge \dot M_{cr},
\end{equation}
and, correspondingly, is
\begin{equation}
\label{mcr2} \dot M_{cr}=\frac{4\pi c}{\kappa} R_{st}.
\end{equation}

The dependence of the Alfven radius  on the accretion rate is such
that the Alfven radius (beyond the capture radius) slightly
decreases with increasing accretion rate as $\dot M^{-1/6}$, while
it decerases below the capture radius as $\dot M^{-2/7}$ and
attains its lowest value for the critical accretion rate $\dot
M_c>\dot M_{cr}$, beyond which it is independent of the external
conditions.

We also note that in the supercritical regime, the pressure of the
accreting plasma increases more slowly (as $R^{-3/2}$) when
approaching the magnetic rotator than the pressure of the
relativistic wind (as $R^{-2}$) ejected by it. This means that in
the supercritical case a cavern  may exist even below the capture
radius.

The estimates presented here, of course, are most suitable for the
case of disk accretion.   In fact, the supercritical regime seems
to emerge most frequently under these conditions. This can be
simply understood. Indeed, the accretion rate is proportional to
the square of the capture radius $\dot M_c\propto R_G^2$. At the
same time, the angular momentum of the captured matter is also
proportional to $R_G^2$. Hence, at high accretion rates the
formation of the disk looks natural.

\subsection{The Effect of the Magnetic Field}

Apparently, the magnetic field of a star becomes significant only
when the stopping radius exceeds the radius of the star,
$R_{st}>R_x$. We take the Alfven radius $R_A$ for $R_{st}$, since
it is the smallest of the two quantities $R_A$ and $R_{Sh}$.
Hence, we can estimate the lowest value of magnetic field of a
star which will influence the flow of matter
\begin{equation}
\label{mf} \begin{array}{l} \mu_{min}= \\ \left\{
\begin{array}{l}
\left.\begin{array}{l}
\left(\frac{\dot M_c v_{\infty}^5 R_x^6}{4G^2M^2_x}\right)^{1/2}, R_A>R_G, \\
\left(\dot M_c\sqrt{GM_x}R_x^{7/2}\right)^{1/2}, R_A\le R_G\\
\end{array}\right\} \dot M_c\le \dot M_{cr},\\
\left(\frac{4\pi c\sqrt{2GM_x}R_x^{9/2}}{\kappa}\right),\quad \dot
M_c>\dot M_{cr},
\end{array}
\right.
\end{array}
\end{equation}

The case $R_A\ge R_G$, $\dot M_c\ge \dot M_{cr}$ is considered
most frequently, and for this case we get the following numerical
estimations
\begin{equation}
\label{est} \mu_{min}=10^{26} R_6^{7/4} \dot M^{1/2}_{17}
m_x^{1/4} \quad \mbox{G cm$^3$},
\end{equation}
or, equivalently,
\begin{equation}
\label{est} B_{min}=10^7 R_6^{5/4} \dot M^{1/2}_{17} m_x^{1/4}
\quad \mbox{G},
\end{equation}

Most presently observed NS have magnetic fields $\sim 10^{12}$ G
and dipole moments $\sim 10^{30}$ G cm$^3$, so the magnetic field
must necessarily be taken into account when considering
interaction of matter with these stars.

\subsection{The Corotation Radius}

The corotation radius is another important characteristics of a
magnetic rotator. Suppose that an accreting plasma penetrates the
light cylinder and is stopped by the magnetic field  at a certain
distance $R_{st}$ given by the balance between the static magnetic
field pressure and the plasma pressure. Suppose that the plasma is
``frozen'' in the rotator's magnetic field. This field will drag
the plasma and force it to rotate rigidly with the angular
velocity of the star. The matter will fall on to the stellar
surface only if its rotational velocity is smaller than the
Keplerian velocity  at the given distance $R_{st}$:

\begin{equation}
\label{rst} \omega R_{st}< \sqrt{GM_x/R_{st}},
\end{equation}

Otherwise, a centrifugal barrier emerges and the rapidly rotating
magnetic field impedes the accretion of matter
\citep{shvartsman1970a,pringle1972a,davidson1973a,lamb1973a,illarionov1975a}.
The latter authors assumed that if $\omega R_{st}\gg
\sqrt{GM_x/R_{st}}$, the magnetic field throws the plasma back
beyond the capture radius. They called this effect the
``propeller'' regime. In fact, matter may not be shed
\citep{lipunov1982a}, but it is important to note that a
stationary accretion  is also not possible.

The corotation radius is thus defined as
\begin{equation}
\label{rst2} R_c=(GM_x/\omega^2)^{1/3}\sim 2.8\times 10^8
m_x^{1/3} (P/1s)^{2/3}\quad\mbox{cm},
\end{equation}
where $P$ is the rotational period of the star.

If $R_{st}<R_c$ , rotation influences the accretion
insignificantly. Otherwise, a stationary accretion  is not
possible for $R_{st}>R_c$.

\subsection{Nomenclature}

The interaction of a magnetic rotator with the surrounding plasma
to a large extent depends on the relation between the four
characteristic radii: the stopping radius, $R_{st}$, the
gravitational capture radius, $R_G$, the light cylinder radius,
$R_l$, and the corotation radius, $R_c$. The difference between
the interaction regimes is so significant that the magnetic
rotators  behave entirely differently in different regimes. Hence,
the classification of the interaction regimes may well mean the
classification of magnetic rotators. The classification notation
and terminology is described below and summarized in Table 1,
based on paper by \citet{lipunov1987a}.

\begin{table*}
\begin{center} \centering \caption{Classification of neutron stars and white dwarfs.}
\begin{tabular}{lllll}
\tableline \tableline
Abbrevi- & Type & Characteristic & Accretion & Well known \\
ation &  & radii relation & rate & observational appearances \\
\tableline
E & Ejector & $R_{st}>R_G$ & $\dot M\le \dot M_{cr}$ & Radio pulsars \\
& & $R_{st}>R_l$ & & \\
P & Propeller & $R_c> R_{st}$ & $\dot M_c\le \dot M_{cr}$ & ? \\
& & $R_{st}\le R_G$ & & \\
& & $R_{st}\le R_l$ & & \\
A & Accretor & $R_{st}\le R_G$ & $\dot M_c\le \dot M_{cr}$ & X-ray pulsars \\
& & $R_{st}\le R_l$ & & \\
G & Georotator & $R_G\le R_{st}$ & $\dot M_c\le \dot M_{cr}$ & ? \\
&  & $R_{st}\le R_c$ & & \\
M & Magnetor & $R_{st}>a$ & $\dot M_c\le \dot M_{cr}$ & AM Her, polars,\\
& & $R_c>a$ ? & & soft gamma repeaters, \\
& & & & anomalous X-ray pulsars \\
SE & Super- & $R_{st}>R_l$ & $\dot M_c>\dot M_{cr}$ & ? \\
& ejector & & & \\
SP & Super- & $R_c<R_{st}$ & $\dot M_c>\dot M_{cr}$ & ? \\
& propeller & $R_{st}\le R_l$ & & \\
SA & Super- & $R_{st}\le R_c$ & $\dot M_c>\dot M_{cr}$ & ? \\
& accretor & $R_{st}\le R_G$ & \\
\tableline
\end{tabular}
\end{center}
\end{table*}

Naturally, not all possible combinations of the characteristic
radii can be realized. For example, the inequality $R_l>R_c$ is
not possible in principle. Furthermore, some combinations require
unrealistically large or small parameters of magnetic rotators.
Under the same intrinsic and external conditions, the same rotator
may gradually pass through several interaction regimes. Such a
process will be referred to as the {\it evolution of a magnetic
rotator}.

We describe the classification by considering an idealized
scenario of evolution of magnetic rotators. Suppose the parameters
$\rho_{\infty}$, $v_{\infty}$ and $\dot M_c$ of the surrounding
medium remain unchanged. We shall also assume for a while a
constancy of the rotator's magnetic moment $\mu$. Let the
potential accretion rate $\dot M_c$ at the beginning be not too
high, so that the reverse effect of radiation pressure can be
neglected, $\dot M_c\le \dot M_{cr}$. We also assume that the star
initially rotates at a high enough speed to provide a powerful
relativistic wind.

{\bf Ejectors (E).} We shall call a magnetic rotator an {\it
ejecting star} (or simply an {\it ejector E}) if the pressure of
the electromagnetic radiation and ejected relativistic particles
is so high that the surrounding matter is swept away beyond the
capture radius or radius of the light cylinder (if $R_l>R_G$).

\begin{equation}
\label{rshe}\mbox{\bf Ejector:}\quad R_{Sh}>\mbox{max}(R_l,R_G),
\end{equation}

It follows from here that $P_m\propto R^{-2}$ while the accretion
pressure within the capture radius is $P_a\propto R^{5/2}$ i.e.
increases more rapidly as we approach an accreting star.
Consequently, the radius of a stable cavern  must exceed the
capture radius \citep{shvartsman1970a}.

It is worth noting that the reverse transition from the propeller
(P) stage to the ejector (E)  stage is non-symmetrical and occurs
at a lower period (see below). This means that to switch a pulsar
on is more difficult than to turn it off. This is due to the fact
that in the case of turning-on of the pulsar the pressures of
plasma and relativistic wind  must be matched at the surface of
the light cylinder, not at the gravitational capture radius. In
fact, the reverse transition occurs under the condition of
equality of the Alfven radius to the radius of the light cylinder
($R_A=R_l$).

It should be emphasized that, as mentioned by
\citep{shvartsman1970a}, relativistic particles can be formed also
at the propeller  stage by a rapidly rotating magnetic field (see
also \citet{kundt1990a}).

{\bf Propellers (P).}  After the ejector stage, the propeller
stage sets in under quite general conditions, when accreting
matter at the Alfven surface is hampered by a rapidly rotating
magnetic field of the magnetic rotator. In this regime the Alfven
radius is greater than the corotation radius, $R_A>R_c$. A finite
magnetic viscosity causes the angular momentum to be transferred
to the accreting matter so that the rotator spins down.  Until
now, the propeller stage is one of the poorly investigated
phenomena. However, it is clear that sooner or later the magnetic
rotator is spin down enough for the rotational effects to be of no
importance any longer, and the accretion stage sets in.

{\bf Accretors (A).} In the accretion stage, the stopping radius
(Alfven radius) must be smaller than the corotation radius
$R_A<R_c$. This is the most thoroughly investigated regime of
interaction of magnetic rotators with accreting plasma. Examples
of such systems span a wide range of bright observational
phenomena from X-ray pulsars,  X-ray bursters,  low-mass X-ray
binaries to most of the cataclysmic variables  and X-ray transient
sources.

{\bf Georotators (G).} Imagine that the star begins rotating so
slowly that it cannot impede the accretion of plasma, i.e. all the
conditions mentioned in the previous paragraph are satisfied.
However, matter still can not fall on to the rotator's surface if
the Alfven radius is larger than the gravitational capture radius
\citep{illarionov1975a,lipunov1982c}. This means that the
attractive gravitational force of the star at the Alfven surface
is not significant. A similar situation occurs in the interaction
of solar wind with Earth's magnetosphere.  The plasma mainly flows
around the Earth's magnetosphere  and recedes to infinity. This
analogy explains the term ``georotator'' used for this stage.
Clearly, a georotator must either have a strong magnetic field or
be embedded in a strongly rarefied medium.

{\bf Magnetors (M).}  When a rotator enters a binary system, it
may happen that its magnetosphere  engulfs the secondary star.
Such a regime was considered by \citet{mitrofanov1977a} for WD in
close binary systems called polars  due to their strongly
polarized emission. In the case of NS, magnetors M may be realized
only under the extreme condition of very close binaries with no
matter within the binary separation.

{\bf Supercritical interaction regimes.} So far, we have assumed
that the luminosity at the stopping surface is lower than the
Eddington limit.  This is fully justified for G and M regimes
since gravitation is not important for them. For types E, P, and
especially A, however, this is not always true. The critical
accretion rate for which the Eddington limit is achieved is
\begin{equation}
\label{mit} \dot M_{cr}=1.5\times 10^{-6} R_8
M_{\odot}\quad\mbox{yr$^{-1}$},
\end{equation}
\noindent where $R_8\equiv R_{st}/10^8$ cm is the stopping radius
(Schwartzman radius  or Alfven radius,  see above).

We stress here that the widely used condition of supercritical
accretion  rate
$$
\dot M\gtrsim 10^{-8} (M/M_{\odot}) M_{\odot}\quad\mbox{yr$^{-1}$}
$$
is valid {\it only} for the case of non-magnetic NS, where
$R_{st}\approx 10$ km coincides with the stellar radius. In
reality, for a NS with a typical magnetic field of
$10^{11}-10^{12}$ G, the Alfven radius reaches $10^7-10^8$ cm, so
much higher accretion rates are required for the supercritical
accretion to set in. The electromagnetic luminosity released at
the NS surface, however, will be restricted by $L_{edd}$, and most
of the liberated energy may be carried away by neutrinos
\citep{basko1975a} (see also section about hyper accretion).

Most of the matter in the dynamic model of supercritical accretion
forms an outflowing flux covering the magnetic rotator by an
opaque shell \citep{shakura1973a}. The following three additional
types are distinguished, depending on the relationship between the
characteristic radii: superejector (SE), superpropeller  (SP) and
superaccretor (SA).

\subsection{A Universal Diagram for Gravimagnetic Rotators}

The classification  given above was based on relations between the
characteristic radii, i.e. quantities which cannot be observed
directly. This drawback can be removed if we note that the light
cylinder radius $R_l$, Shvartzman radius $R_{Sh}$ and corotation
radius $R_c$ are functions of the well-observed quantity,
rotational period of the magnetor $p$. Hence, the above
classification can be reformulated in the form of inequalities for
the rotational period of a magnetic rotator.

One can introduce two critical periods $p_E$ and $p_A$ such that
their relationship with period $p$ of a magnetic rotator specifies
the rotator's type:

\begin{equation}
\begin{array}{l}
p<p_E,\quad\to\quad\mbox{E or SE}, \\
p_E\le p<p_A,\quad\to\quad\mbox{P or SP}, \\
p>p_A,\quad\to\quad\mbox{A, SA, G or M},
\end{array}
\end{equation}

The values of $p_E$ and $p_A$ can be determined from Table 2 which
defines the basic nomenclature, and are functions of the
parameters $v_{\infty}$, $\dot M_c$, $\mu$ and $M_x$. The
parameters $p$ and $\mu$ characterize the electromagnetic
interaction, while $\dot M_c$ describes the gravitational
interaction. Instead of $\dot M_c$ we introduce the potential
accretion luminosity $L$

\begin{equation}
\label{l} L\equiv \dot M_c\frac{GM_x}{R_x},
\end{equation}

\begin{table*}
\begin{center} \caption{Parameter of the evolution equation of a magnetic rotators. }
\begin{tabular}{lllllll}
\tableline
Parameter & Regime \\
& E, SE & P, SP & A & SA & G & M \\
\tableline
\\
$\dot M$ & 0 & 0 & $\dot M_c$ & $\dot M_c(R_A/R_s)$ & 0 & $\dot M_c$ \\
$\kappa_t$ & $\sim 2/3$ & $\lesssim 1/3$ & $\sim 1/3$ & $\sim 1/3$ & $\sim 1/3$ & $\sim 1/3$ \\
$R_t$ & $R_l$ & $R_m$ & $R_c$ & $R_c$ & $R_A$ & a \\
\tableline
\end{tabular}
\end{center}
\end{table*}

The physical sense of the potential luminosity is quite clear: the
accreting star would be observed to have this luminosity if the
matter formally falling on the gravitational capture cross-section
were to reach its surface.

Approximate expressions for critical periods \citep{lipunov1992a}
are:
\begin{equation}
\label{pe} p_E=\left\{\begin{array}{l}
0.42 v_7^{-1/4} \mu_{30}^{1/2} L_{38}^{-1/4} \quad \mbox{s}, \hspace{16pt} (\alpha) \\
1.8 v_7^{-5/6} m^{1/3} \mu_{30}^{1/3} L_{38}^{-1/6} \quad \mbox{s}, (\beta) \\
1.4\cdot 10^{-2} m^{-1/9} \mu_{30}^{4/9} \quad \mbox{s},
\hspace{44pt} (\gamma)
\end{array}\right.
\end{equation}

\noindent We assume the next indication in this formula:
($\alpha$), $\dot M_c\le \dot M_{cr}$, $p\le p_{GL}$; ($\beta$),
$\dot M_c\le \dot M_{cr}$, $p>p_{GL}$; ($\gamma$), $\dot M_c >
\dot M_{cr}$.

\begin{equation}
\label{pa} p_A=\left\{\begin{array}{l}
400 v_7^{-5/4} \mu_{30}^{1/2} L_{38}^{-1/4} ,  \quad \mbox{s} ,\quad (\alpha) \\
1.2 m^{-5/7} \mu_{30}^{6/7} L_{38}^{-3/7} ,  \quad \mbox{s} , \quad (\beta) \\
0.17 m^{-2/3} \mu_{30}^{2/3} , \quad \mbox{s} , \hspace{9pt}\qquad
(\gamma)
\end{array}\right.
\end{equation}

\noindent We assume the next indication in this formula:
($\alpha$), $\dot M_c\le \dot M_{cr} \mbox{ and } R_A>R_G$;
($\beta$), $\dot M_c\le \dot M_{cr} \mbox{ and } R_A\le R_G$;
($\gamma$), $\dot M_c>\dot M_{cr}$.

Here a new critical period $p_{GL}$ was introduced from the
condition $R_G=R_l$:
\begin{equation}
\label{rgrl} p_{GL}=\frac{4\pi GM_x}{v_{\infty}^2 c}\approx 500
m_x v_7^{-2} \quad\mbox{s},
\end{equation}

Treating the rotator's magnetic dipole moment $\mu$ and $M_x$ as
parameters, we find that an overwhelming majority of the
magnetor's stages can be shown on a ``p-L'' diagram
\citep{lipunov1982a}. The quantity $L$ also proves to be
convenient because it can be observed directly at the accretion
stage.

\subsection{The Gravimagnetic Parameter}

By expecting the expression for the stopping radius  in the
subcritical regime ($\dot M_c\le \dot M_{cr}$) one can note that
the magnetic dipole moment $\mu$ and the accretion rate $\dot M_c$
always appear in the same combination,

\begin{equation}
\label{gpp} y=\frac{\dot M_c}{\mu^2},
\end{equation}

\noindent as was noticed by \citet{davies1981a}. The parameter $y$
characterizes the ratio between the gravitational and magnetic
``properties'' of a star and will, therefore, be called the {\it
gravimagnetic parameter}. Two magnetic rotators having quite
different magnetic fields, subjected to different external
conditions but with identical gravimagnetic parameters, have
similar magnetospheres, as long as the accretion rate is quite low
($\dot M_c\le \dot M_{cr}$). Otherwise, the flux of matter near
the stopping radius no longer depends on the accretion rate at a
large distance.

In fact, the number of independent parameters can be further
reduced (see e.g. \citet{lipunov1992a}) by introducing the
parameter

\begin{equation}
\label{gpp} Y=\frac{\dot M_c v_{\infty}}{\mu^2},
\end{equation}

Plotting the rotator's period $p$ versus $Y$-parameter  we can
draw a somewhat less obvious but more general classification
diagram  than the ``p-L'' diagram discussed above. This permits us
to show on a single plot the rotators with key parameters $\dot
M_c$, $\mu$ and $v_{\infty}$ spanning a very wide range.

In the case of supercritical accretion,  another characteristic
combination is found in all the expressions:

\begin{equation}
\label{gpp} Y_s=\frac{\kappa \mu^2}{v_{\infty}},
\end{equation}

In analog to the subcritical ``p-Y'' diagram, a supercritical
``p-Y$_s$'' diagram can be drawn.

\section{Evolution of Magnetic Rotators}

The evolution of a magnetic rotator, which determines its
observational manifestations, involves the slow changing of the
regimes of its interaction with the surrounding medium. Such an
approach to the evolution was developed in the 1970s by
\citet{shvartsman1970a,biskogan1976a,illarionov1975a,shakura1975a,wickram1975a},
\\ \citet{heuvel1977a} and others. Three stages were mostly
considered in these papers: ejector, propeller  and accretor.  All
these stages can be described by a unified evolutionary equation.

\subsection{The evolution equation}

Analysis of the nature of interaction of a magnetized star with
the surrounding plasma allows us to write an approximate evolution
equation for the angular momentum of a magnetic rotator in the
general form \citep{lipunov1982a}:

\begin{equation}\label{ksuevol}
\frac{\textit{d}I\omega}{\textit{d}t} = \dot{M} k_{su} -
\kappa_t\frac{\mu^2}{R^3_t} ,
\end{equation}

\noindent where $k_{su}$ is a specific angular momentum applied by
the accretion matter to the rotator. This quantity is given by

\begin{equation}
\label{ksu} k_{su}=\left\{\begin{array}{l}
(GM_x R_d)^{1/2} , \quad \mbox{Keplerian disk accretion,}\\
\eta_t \Omega R_G^2, \quad \mbox{wind accretion in a binary,}\\
\sim 0 , \quad \mbox{a single magnetic rotator.}
\end{array} \right.
\end{equation}

\noindent where $R_d$ is the radius of the inner disk edge,
$\Omega$ is the rotational frequency of the binary system, and
$\eta_t\approx 1/4$ \citep{illarionov1975a}. The values of
dimensionless factor $\kappa_t$, characteristic radius $R_t$ and
the accretion rate $\dot M$ in different regimes are presented in
Table 2.

The evolution equation (\ref{ksuevol}) is approximate. In
practice, the situation with propellers  and superpropellers  is
not yet clear. In Table 2 $R_m$ is the size of a magnetosphere
whose value at the propeller stage is not known accurately and
which may differ significantly from the standard expressions for
the Alfven radius.

\subsection{The equilibrium period}

The evolution equation presented above indicates that an accreting
compact star must endeavor to attain an equilibrium state in which
the resultant torque vanishes \citep{davidson1973a,lipunov1976a}.
This hypothesis is confirmed by observations of X-ray pulsars.

By equating the right-hand side of equation (\ref{ksu}) to zero,
we obtain the equilibrium period:

\begin{equation}
\label{prt} \begin{array}{l}
p_{eq}\approx 7.8\pi \sqrt{\kappa_t/\epsilon^2} (GM_x)^{-5/7} y^{-3/7}\quad\mbox{s},\quad (\alpha)\\
p_{eq}=\sqrt{A/B_w} L_{37}^{-1/2} T_{10}^{-1/6}\quad\mbox{s},\quad
(\beta)
\end{array}
\end{equation}

\noindent where $A\approx 5\times 10^{-4 } (3\kappa_t)\mu_{30}^2
I_{45}^{-1} m_x^{-1}\quad\mbox{s yr$^{-1}$}$, and $B_w\approx
5.2\times 10^{-6} R_6^2 m_0^{2/3} / (10^{2/3}m_x^2)
I_{45}^{-1}\dot M_{-6}\eta \quad\mbox{s yr$^{-1}$}$,
$L_{37}=L/10^{37}$, $T_{10}=T/10$ days; ($\alpha$), disk
accretion; ($\beta$), quasi-spherical accretion.

Alternatively:
\begin{equation}
\label{prt} \begin{array}{l}
p_{eq}\approx 1.0L_{37}^{-3/7}\mu_{30}^{6/7}\quad\mbox{s},\quad \mbox{disk},\\
p_{eq}=10\eta^{-1/2}_{k}\dot M_{-6}^{-1/2}\times \\
\times(m_0^{2/3}/(10^{2/3}
m_x^2))^{-1/2}\times \\
\times L_{37}^{-1}T_{10}^{-1/6}\mu_{30}\quad\mbox{s},\quad
\mbox{stellar wind},
\end{array}
\end{equation}

Let us turn to the case of disk accretion.  The above model of the
spin-up and spin-down torques  possesses an unexpected property.
The equilibrium period  obtained by setting the torque to zero is
connected with the critical period $p_A$ through a dimensionless
factor:

\begin{equation}
\label{peqa} p_{eq}(A)=2^{3/4}
\frac{\kappa_t^{1/2}}{\varepsilon^{7/4}}p_A,
\end{equation}

The parameters $\kappa_t$ and $\varepsilon$ must be such that
$p_{eq}>p_A$. Since $\kappa_t\approx\varepsilon\approx 1$, the
equilibrium period in the case of disk accretion is close to the
critical period, $p_A$, separating accretion stage $A$ and the
propeller stage $P$. In the case of the supercritical accretion
the equilibrium period is determined by formula

\citep{lipunov1982b}:
\begin{equation}
\label{l2} p_{eq}(SA)\simeq 0.17 \mu_{30}^{2/3}
m_x^{-1/9}\quad\mbox{s},
\end{equation}

\subsection{Evolutionary Tracks}

The evolution of NS in binaries must be studied in conjunction
with the evolution of normal stars. This problem was discussed
qualitatively by \citet{biskogan1976a,heuvel1977a},
\\ \citet{lipunov1982a} and other. We begin with the qualitative analysis
presented in the latter of these paper.

The most convenient method of analysis of NS evolution is using
the ''p-L'' diagram. It should be recalled that $L$ is just the
potential accretion luminosity   of the NS. This quantity is equal
to the real luminosity only at the accretion stage.

In Figure 1 we show the evolutionary tracks of a NS. As a rule, a
NS in a binary is born when the companion star belongs to the main
sequence (loop-like track). During the first $10^5-10^7$ years,
the NS is at the ejector  stage, and usually it is not seen as a
radiopulsar since its pulse radiation is absorbed in the stellar
wind  of the normal star. The period of the NS increases in
accordance with the magnetic dipole losses.  After this, the
matter penetrates into the light cylinder and the NS passes first
into the propeller  stage and then into the accretor stage. By
this time, the normal star leaves the main sequence and the
stellar wind strongly increases. This results in the emergence of
a bright X-ray pulsar. The period of the NS stabilizes around its
equilibrium value. Finally, the normal star fills the Roche lobe
and the accretion rate suddenly increases; the NS moves first to
the right and then vertically downward in the ''p-L'' diagram. In
other words, the NS enters the supercritical stage SA
(superaccretor)  and its spin period tends to a new equilibrium
value  (see equation (\ref{l2})).

\begin{figure*}
\epsscale{1.0} \plotone{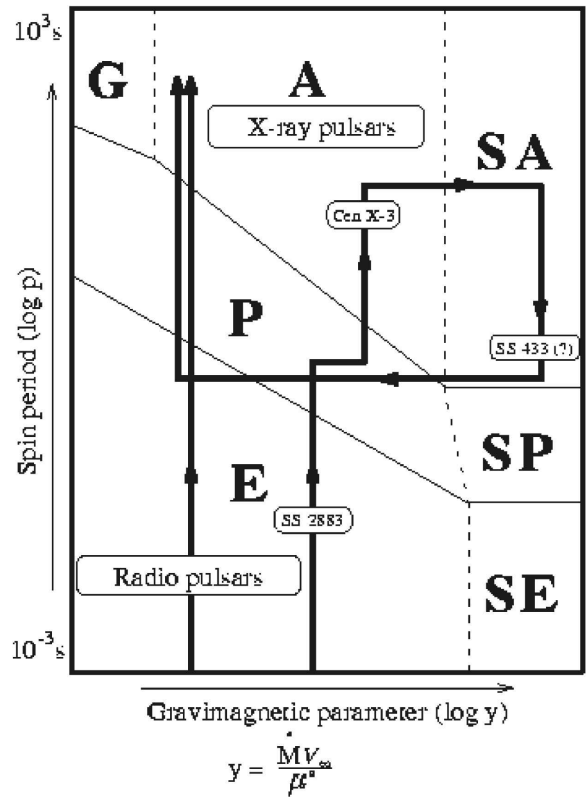} \figcaption{Tracks of NS on the
period (p) - gravimagnetic parameter (Y) diagram: track of a
single NS (vertical line) and of a NS in a binary system (looped
line). For the second track, possible observational appearances of
the NS are indicated.} \label{figep}
\end{figure*}

After the mass exchange, only the helium core of the normal star
is left (a WR star in the case of massive stars), the system
becomes detached and the NS returns back to the propeller  or
ejector state. Accretion is still hampered by rapid NS rotation.
This is probably the reason underlying the absence of X-ray
pulsars  in pairs with Wolf-Rayet stars \citep{lipunov1982c}.
Since the helium star evolves on a rather short time-scale
($\approx 10^5$ yr), the NS does not have time to spindown
considerably: after explosion of the normal star, the system can
be disrupted leaving the old NS as an ejector, i.e. as a
high-velocity radio pulsar.

The ``loop-shaped'' track discussed above can be written in the
form:

\begin{itemize}

\item I+E $\to$ I+P $\to$ II+P $\to$ II+A $\to$ III+SA $\to$ IV+P
$\to$ E+E (recycled pulsar) $\to$ \dots

\item I+E $\to$ I+P $\to$ II+A $\to$ III+SA $\to$ IV+E $\to$
(recycled ejector) $\to$ IV+P $\to$ E+E (recycled pulsar) $\to$
\dots

\item Another version of the evolutionary track of a NS formed in
the process of mass exchange within a binary system is: \\
III+SE $\to$ III+SP $\to$ IV+P $\to$ E+E $\to$ \dots

\end{itemize}

The overall lifetime of a NS in a binary system depends on the
lifetime of the normal star and on the parameters of the binary
system. However, the number of transitions from one stage to
another during the time the NS is in the binary is proportional to
the magnetic field  strength of the NS.

Figure 2 demonstrates the effect of NS magnetic field decay (track
(a) with and (b) without magnetic field decay). The first track
illustrates the common path which results in the production of a
typical millisecond pulsar.

\begin{figure*}
\epsscale{1.0} \plotone{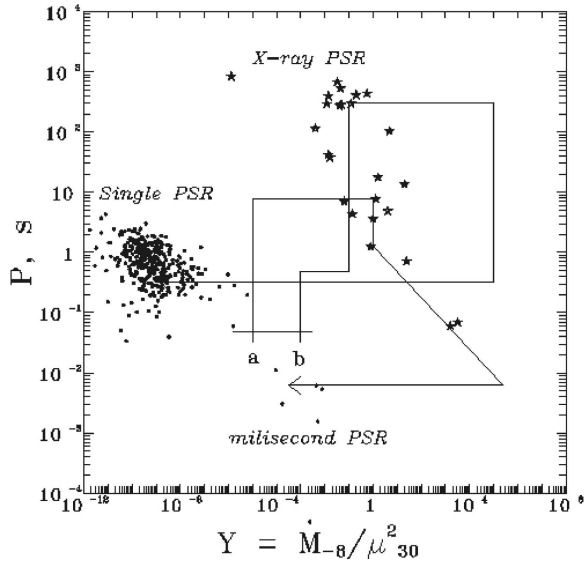} \figcaption{The
period-gravimagnetic parameter diagram for NS in binary systems.
(a) with NS magnetic field decay  (the oblique part of the track
corresponds to ``movement'' of the accreting NS along the
so-called ``spin-up''  line), (b) a typical track of a NS without
field decay in a massive binary system.} \label{figep}
\end{figure*}

\subsection{Evolution of Magnetic Rotators in Non-circular
Orbits}

So far we have considered evolution of a magnetic rotator related
to single rotators or those entering binary system with circular
orbits. This approximation was appropriate for the gross analysis
of binary evolutionary scenario performed by
\citet{kornilov1983a,kornilov1983b}. This approximation is further
justified by the fact that the tidal interaction in close binaries
leads to orbital circularization in a short time \citep{hut1981a}.
However, the more general case of a binary with eccentric orbit
must be considered for further analysis. It is especially
important because many of the currently observed X-ray pulsars, as
well as radio pulsars with massive companions PSR B1259-63
\citep{johnston1992a} and PSR B0042-73 \citep{kaspi1994a}, are in
highly eccentric orbits around massive companions. Previously,
such studies have been performed by
\citet{gnus1985a,prokhorov1987a}.

Orbital eccentricity necessarily emerges after the first supernova
explosion  and mass expulsion from the binary system. In massive
binaries with long orbital periods $\gtrsim 10$ days, the
eccentricity may be well conserved until the second episode of
mass exchange \citep{hut1981a}. Here, we concentrate on the
evolutionary consequences of eccentricity.

\subsection{Mixing types of E-P-A binary systems with non-zero orbital
eccentricity}

The impact of eccentricity on the observed properties of X-ray
pulsars has been considered in many papers
\citep{amnuel1971a,shakura1973a,pacini1975a,lipunov1976a}. The
most important consequence of orbital eccentricity for the
evolution of rotators can be understood without detailed
calculations, and suggests the existence of two different types of
binary systems separated by a critical eccentricity, $e_{cr}$
\citep{gnus1985a}.

Consider an ideal situation when a rotator enters a binary system
with some eccentricity. The normal star (no matter how) supplies
matter to the compact magnetized rotator. We assume that all the
parameters of the binary system (binary separation, eccentricity,
masses, accretion rate, etc.) are stationary and unchanged. Then a
critical eccentricity $e_{cr}$ appears such that at $e>e_{cr}$ the
rotator is not able to reach the accretion state in principle. Let
the rotator be rapid enough initially to be at the ejector (E)
state. With other parameters constant, the evolution of such a
star is determined only by its spindown.  The star will gradually
spindown to such a state that when passing close to the periastron
where the density of the surrounding matter is higher, the pulsar
will ``choke'' with plasma and pass into the propeller  regime.
Therefore, for a small part of its life the rotator will be in a
mixed EP-state, being in the propeller state at periastron and at
the ejector state close to apastron. The subsequent spindown of
the rotator leads most probably to the propeller state along the
entire orbit. This is due to the fact that the pressure of matter
penetrating the light cylinder $R_l$ increases faster than that
caused by relativistic wind and radiation, as first noted by
\citet{shvartsman1971a}. So it proves to be much harder for the
rotator to pass from the P state to the E state than from E to P
state (see the following section).

The rotator will spindown  ultimately to some period, $p_A$, at
which accretion will be possible during the periastron passage.
Accretion, in contrast, will lead to a spin-up  of the rotator, so
that it reaches some average equilibrium state characterized by an
equilibrium period $p_{eq}$ defined by the balance of accelerating
and decelerating torques averaged over the orbital period. If the
eccentricity was zero, the rotator would be in the accretion state
all the time. By increasing the eccentricity and keeping the
periastron separation between the stars unchanged, we increase the
contribution of the decelerating torque over the orbital period
and thus decrease $p_{eq}$. At some ultimate large enough
eccentricity $e_{cr}$ the equilibrium period will be less than the
critical period $p_A$ permitting the transition from the propeller
state to the accretion state at apastron to occur. The rotational
torque applied to the rotator, averaged over orbital period,
vanishes, and in this sense the equilibrium state is achieved, but
the rotator periodically passes from the propeller state to the
accretion state.

Thus, X-ray pulsars with unreachable full-orbit accretion state
must exist. This means that from the observational point of view
such binaries will be observed as transient  X-ray sources with
stationary parameters for the normal component.

Typically, the evolutionary track of a rotator in an eccentric
binary is

\begin{itemize}

\item E $\to$ PE $\to$ P $\to$ AP, $\quad$ $e>e_{cr}$

\item E $\to$ PE $\to$ P $\to$ AP $\to$ A,$\quad$ $e<e_{cr}$

\end{itemize}
this may be the principal formation channel of transient  X-ray sources.

\subsection{Ejector-propeller hysteresis}

As mentioned earlier, the transition of the rotator from the
ejector state to the propeller state is not symmetrical. Here we
consider this effect in more detail. In terms of our approach, we
must study the dependence of $R_{st}$ on $\dot M_c$. To find
$R_{st}$, we must match the ram pressure of the accreting plasma
with that caused by the relativistic wind or by the magnetosphere
of the rotator. This dependence $R_{st}(\dot M_c)$  will be
substantially different for rapidly ($R_l<R_G$) and slowly
($R_l>R_G$) rotating stars (see Figure 3). One can see that in the
case of a fast rotator, an interval of $\dot M_c$ appears where
three different values of $R_{st}$ are possible, the upper value
$R_1$ corresponding to the ejector state and the bottom value
$R_3$ to the propeller state; the intermediate value $R_2$ is
unstable. This means that the rotator's state is not determined
solely by the value of $\dot M_c$ , but also depends on previous
behavior of this value.

\begin{figure*}
\epsscale{1.0} \plotone{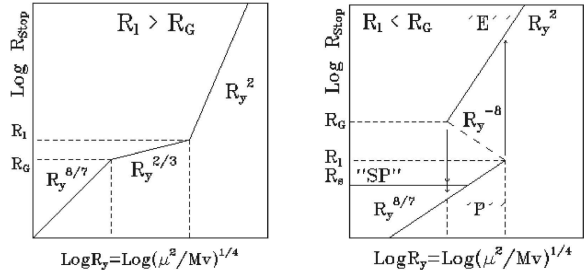} \figcaption{Dependence of the
stopping radius $R_{st}$ on the modified gravimagnetic radius
$R_y$ for two possible relations between the light cylinder radius
$R_l$ and the gravitation capture radius $R_G$: $R_l>R_G$
(left-hand panel) and $R_l<R_G$ (right-hand panel; here the
ejector-propeller hysteresis becomes possible).} \label{figep}
\end{figure*}

Now consider a periodic changing of $\dot M_c$  caused, for
example, by the rotator's motion along an eccentric orbit, and
large enough for the rotator to transit from the ejector  state to
the propeller state and vice versa. Initially, the rotator is in
the ejector state. By approaching the normal star, the accretion
rate $\dot M_c$ increases and reaches a critical value $\dot
M_{EP}$, where the equilibrium points $R_1$ (stable point
corresponding to the ejector state) and $R_2$ (unstable) approach
$R_G$ (upper kink), where they merge (see Figure 2). After that
only one equilibrium point remains in the system, the stopping
radius $R_{st}$ jumps from $\approx R_G$ down to $R_3<R_l$, and
the rotator changes to the propeller state.

As $\dot M_c$ decreases further along the orbit and reaches the
critical value $\dot M_{EP}$ once again, the reverse transition
from propeller to ejector does not occur. The transition only
occurs when $\dot M_c$ reaches another critical value, $\dot
M_{PE}<\dot M_{EP}$, where the unstable point $R_2$ meets the
stable propeller point $R_3$, and the stopping radius $R_{st}$
jumps from $\sim R_l$ up to $R_1>R_G$. It should be noted that for
fast enough rotators, a situation is possible when the step down
from the ejector state occurs in such a manner that the stopping
radius $R_{st}<R_c$ and the rotator passes directly to the
accretion state.  The reverse transition always passes through the
propeller stage: A $\to$ P $\to$ E. In principle, transitions from
the ejector state to supercritical states SP or SA are also
possible \citep{prokhorov1987a}. In the case of slow rotators
($R_l>R_B$)  ), the ``E-P'' hysteresis is not possible, and
transitions between these states are symmetrical.

\subsection{E-P transitions for different orbits}

Orbital motion of the rotator around the normal companion in an
eccentric binary draws a horizontal line on the ''p-Y'' diagram,
with the beginning at a point corresponding to $\dot M_c(a_p)$,
and the end at a point corresponding to $\dot M_c(a_a)$ (here
$a_p$ and $a_a$ are the periasrton and apastron distances,
respectively). The length of this segment is determined by the
eccentricity. Since $Y\propto \dot M_c$ , the rotator moves along
this segment from left to right and back as it revolves from the
apastron to the periastron. At each successive orbital period,
this line slowly drifts up to larger periods. The evolution of
this system is thus determined by the order the critical lines on
the diagram are crossed by this ``line''. It is seen from the
''p-Y'' diagram that the regions with and without hysteresis are
separated by a certain value of the parameter $Y=Y_k$. Since
$Y\propto \dot M_c\propto 1/r^2$ , four different situations are
possible depending on the relationship of the binary orbital
separation a with critical value $a_{cr}$, corresponding to $Y_k$
(see e.g. \citet{lipunov1992a,lipunov1996b}).

\begin{enumerate}

\item $\bf a_p>a_{cr}$. In this case no hysteresis occurs and
transitions E--P and the reverse take place in symmetrical points
of the orbit. The rotator passes the following sequence of stages:
E $\to$ EP $\to$ P $\to$ \dots  Here EP means a mixed state of the
rotator at which it is in the ejector state during one part of the
orbit, and in the propeller state during the rest of the orbital
cycle.

\item $\bf a_a>a_{cr}>a_p$. In this case, the hysteresis occurs at
the beginning of the mixed EP-state (state EP$_h$), but as the
rotator slows down the hysteresis gradually decays and disappears.
E-P transition for this system is: E $\to$ EP$_h$ $\to$ EP $\to$ P
$\to$ \dots

\item $\bf a_a<a_{cr}$. The hysteresis is possible in principle,
but the shape of the transition depends on the eccentricity.
Suppose a pulsar has spun down so much that the first transition
from the ejector to the propeller occurred at the periastron. If
the eccentricity were smal $e<e_{cr}$, (do not confuse this
$e_{cr}$ with the critical eccentricity introduced in the previous
section) the reverse transition to the ejector state would not
occur even at the apastron, and the evolutionary path would be E
$\to$ P $\to$ \dots

\item If $e>e_{cr}$ , the track is E $\to$ EP$_h$ $\to$ P $\to$
\dots It should be noted that just after the first EP transition
(as well as before the last), the system spends a finite time in
the E and P states at every revolution.

\end{enumerate}

The value of $e_{cr}$ can be expressed through the orbital
parameters as
\begin{equation}
\label{bec} e_{cr}\simeq
\frac{a_{cr}^{1/7}-a_a^{1/7}}{a_{cr}^{1/7}+a_a^{1/7}},
\end{equation}

To conclude, we note that the hysteresis during the
ejector-propeller transition may be possible for single radio
pulsars also.  For example, when the pulsar moves through a dense
cloud of interstellar plasma, the pulses can be absorbed. The
radio pulsar turns on again when it comes out from the cloud. The
hysteresis amplitude for single pulsars can be high enough because
of small relative velocities of the interstellar gas and the
pulsar, so that $R_G\gg R_l$.

\section{Summary}

The ``Scenario Machine'' is the numerical code for theoretical
investigations of statistical properties of binary stars, i.e.
this is population synthesis code \citep{lipunov1996b}. It
includes the evolution of normal stars and the evolution of their
compact remnants. This is especially important for studies of the
neutron stars \citep{lipunov1992a}. We always include the most
important observational discoveries and theoretical estimations
into our code.


\begin{thebibliography}{99}

\bibitem[Abt (1983)]{abt1983a} Abt H.A., 1983, Annual review of astronomy and astrophysics, v. 21, p. 343

\bibitem[Allen (1973)]{allen1973a} Allen C.W., 1973, University of London, Athlone Press, 3rd ed.

\bibitem[Amnuel' \& Guseinov (1968)]{amnuel1968a} Amnuel' P.R., Guseinov O.H., 1968, Izv. Akad. Nauk Az. SSR, 3, 70

\bibitem[Amnuel' \& Guseinov (1971)]{amnuel1971a} Amnuel' P.R., Guseinov
O.Kh., 1971, Soviet Astronomy, v. 15, p. 218

\bibitem[Basko \& Sunyaev (1975)]{basko1975a} Basko M.M., Sunyaev R.A., 1975, A\&A, 42, 311

\bibitem[Bisnovatyi-Kogan \& Komberg (1976)]{biskogan1976a} Bisnovatyi-Kogan G.S., Komberg
B.V., 1976, Soviet Astronomy, v. 19, p. 279

\bibitem[Boersma (1961)]{boersma1961a} Boersma J., 1961, Bulletin of the Astronomical Institutes of the Netherlands, v. 15, p. 291

\bibitem[Bogomazov et al. (2005)]{bogomazov2005a} Bogomazov A.I.,
Abubekerov M.K., Lipunov V.M., Cherepashchuk A.M., 2005, Astron.
Reports, 49, 295

\bibitem[Bondi (1952)]{bondi1952a} Bondi H., 1952, MNRAS, 112, 195

\bibitem[Bondi \& Hole (1944)]{bondi1944a} Bondi H., Hole F., 1944, MNRAS, 104, 273

\bibitem[Chanmugam (1992)]{chanmugam1992a} Chanmugam G., 19992, Annual review of astronomy and astrophysics, v. 30, p. 143

\bibitem[Cherepashchuk et al. (1984)]{cherepashchuk1984a} Cherepashchuk A.M., Khaliullin Kh., Eaton
J.A., 1984, ApJ, v. 281, p. 774

\bibitem[Cherepashchuk et al. (1996)]{cherepashchuk1996a}
Cherepashchuk A.M., Katysheva N.A., Khurzina T.S., Shugarov S.Yu.,
1996, Highly Evolved Close Binary Stars: Catalog, Gordon \& Breach
Publ., The Netherlands

\bibitem[Chevalier (1993)]{chevalier1993a} Chevalier R.A., 1993, ApJ, 411, L33

\bibitem[Davidson \& Ostriker (1973)]{davidson1973a} Davidson K., Ostriker
J.P., 1973, ApJ, v. 179, pp. 585

\bibitem[Davies \& Pringle (1981)]{davies1981a} Davies R.E., Pringle J.E., 1981, MNRAS, 196, 209

\bibitem[Delgado \& Thomas (1981)]{delgado1981a} Delgado A.J., Thomas
H.-C., 1981, A\&A, v. 96, p. 142

\bibitem[Eggleton (1983)]{eggleton1983a} Eggleton P.P., 1983, ApJ, v. 268, p. 368

\bibitem[Garcia-Berro \& Iben (1994)]{gb1994} Garcia-Berro E., Iben
I., 1994, ApJ, v. 434, p. 306

\bibitem[Gnusareva \& Lipunov (1985)]{gnus1985a} Gnusareva V.S., Lipunov V.M., 1985, Soviet Astronomy, v.29, p. 645

\bibitem[Hut (1981)]{hut1981a} Hut P., 1981, A\&A, 99, 126

\bibitem[Iben \& Tutukov (1985)]{iben1985a} Iben I. Jr., Tutukov A. V., 1985, ApJ, 58, 661-710

\bibitem[Iben \& Tutukov (1987)]{iben1987a} Iben Icko Jr., Tutukov Alexander V., 1987, ApJ, 313, 727-742

\bibitem[Illarionov \& Sunyaev (1975)]{illarionov1975a} Illarionov A.F., Sunyaev
R.A., 1975, A\&A, v. 39, p. 185

\bibitem[van den Heuvel (1983)]{heuvel1983a} van den Heuvel
E.P.J., 1983, Accretion-driven stellar X-ray sources, p. 303

\bibitem[van den Heuvel \& Heise (1972)]{heuvel1972a} van den Heuvel E.P.J., Heise
J., 1972, Nature Physical Science, v. 239, p. 67

\bibitem[de Jager (1980)]{jager1980a} de Jager C., 1980, The Brightest Stars, Reidel,
Dordrecht

\bibitem[de Jager et al. (1988)]{jager1988a} de Jager C., Nieuwenhuijzen H., van der Hucht
K.A., 1988, A\&AS, v. 72, p. 259

\bibitem[Johnston et al.(1992)]{johnston1992a} Johnston S., Manchester R.N., Lyne A.G., et al., 1992, ApJ, 387, L37

\bibitem[Joss \& Rappaport (1983)]{joss1983a} Joss P.C., Rappaport
S., 1983, ApJL, 270, 73

\bibitem[Kalogera \& Webbink (1998)]{kalogera1998a} Kalogera V.,
Webbink R.F., 1998, ApJ, 493, 351

\bibitem[\protect\citeauthoryear{Karpov \& Lipunov}{2001}]{karpov} Karpov S.V., and Lipunov V.M., 2001, Astron. Letters, 27, 10, 645-647

\bibitem[Kaspi et al.(1994)]{kaspi1994a} Kaspi V.M., Johnston S., Bell J.F., et al., 1994, ApJ, 423, L43

\bibitem[Kawaler (1988)]{kawaler1988a} Kawaler S.D., 1988, ApJ,
333, 236

\bibitem[Kippenhahn \& Weigert (1967)]{kippenhahn1967a} Kippenhahn R., Weigert
A., 1967, Zeitschrift fur Astrophysik, v. 65, p. 251

\bibitem[Kolb et al. (1998)]{kolb1998a} Kolb U., King A.R., Ritter
H., 1998, MNRAS, 298, L29

\bibitem[Kornilov \& Lipunov (1983a)]{kornilov1983a} Kornilov V.G., Lipunov
V.M., 1983, Soviet Astronomy, v. 27, p. 163

\bibitem[Kornilov \& Lipunov (1983b)]{kornilov1983b} Kornilov V.G., Lipunov
V.M., 1983, Soviet Astronomy, v. 27, p. 334

\bibitem[\protect\citeauthoryear{Karpov \& Lipunov}{2001}]{karpov} Karpov S.V., and Lipunov V.M., Astron. Letters, 2001, 27, 10, 645-647

\bibitem[Krajcheva et al. (1981)]{krajcheva1981}
Krajcheva Z.T., Popova E.I., Tutukov A.V., Yungelson L.R., 1981,
SvA Lett., 7, 269

\bibitem[Kudritzki \& Reimers (1978)]{kudritzki1978a} Kudritzki B.P., Reimers D., 1978, A\&A,
70, 227

\bibitem[Kundt (1990)]{kundt1990a} Kundt W., 1990, In Neutron Stars and their Birth Events, ed. W.Kundt, Kluwer Academic Publishers, Dordrecht, p. 1

\bibitem[Kurucz (1991)]{kurucz1991a} Kurucz R.L., New Opacity Calculations, 1991, In Stellar Atmospheres: Beyond Classical Models, Proceedings of the Advanced Research Workshop, Trieste, Italy, Dordrecht, D. Reidel Publishing Co., p.441

\bibitem[Lamb (1973)]{lamb1973a} Lamb F.K., Pathick C.J., Pines D., 1973, ApJ, 184, 271

\bibitem[Lamers (1981)]{lamers1981a} Lamers H. J. G. L. M., 1981, ApJ, 1, 245, 593-608

\bibitem[Landau \& Lifshiz (1971)]{landau1971a} Landau L.D., Lifshiz E.M., 1971, Classical Theory of Fields, Addisson-Wesley, Reading, Massachusetts and Pergamon Press, London

\bibitem[Landre et al. (1990)]{landre1990a} Landre V., Prantzos N., Aguer P., Bogaert G., Lefebvre A., Thibaud
J.P., 1990, A\&A, v. 240, p. 85

\bibitem[Lipunov (1982a)]{lipunov1982a} Lipunov V.M., 1982a, Ap\&SS,
82, 343

\bibitem[Lipunov (1982b)]{lipunov1982b} Lipunov V.M., 1982b, SvA, 26, 54

\bibitem[Lipunov (1982c)]{lipunov1982c} Lipunov V.M., 1982c, SvAL, 8, 194

\bibitem[Lipunov (1982d)]{lipunov1982d} Lipunov V.M., 1982, Soviet
Astronomy, 26, 537

\bibitem[Lipunov (1984)]{lipunov1984a} Lipunov V.M., 1984, Advances in Space
Research, v. 3., no. 10-12, p. 323

\bibitem[Lipunov (1987)]{lipunov1987a} Lipunov V.M., 1987, Ap\&SS, 132, 1

\bibitem[Lipunov (1992)]{lipunov1992a} Lipunov V.M., 1992, Astrophysics of Neutron Stars, Springer-Verlag, Berlin -
Heidelberg - New York, Astronomy and Astrophysics Library, 322

\bibitem[Lipunov (2006)]{lipunov2006a} Lipunov V.M, 2006, IAU proseedings,
Populations of High Energy Sources in Galaxies Proceedings of the
230th Symposium of the International Astronomical Union, Edited by
E.J.A. Meurs, G. Fabbiano, Cambridge University Press, 2006, p.
391 iu Q.Z., van Paradijs J., van den Heuvel E.P.J., 2000, A\&A,
368, 1021

\bibitem[Lipunov et al. (1996a)]{lipunov1996a} Lipunov V.M., Ozernoy L.M., Popov S.B., Postnov K.A.,
Prokhorov M.E., 1996a, ApJ, 466, 234

\bibitem[Lipunov \& Postnov (1988)]{lipunov1988a} Lipunov V.M., Postnov
K.A., 1988, Ap\&SS, v. 145, no. 1, p. 1-45.

\bibitem[Lipunov et al. (1996b)]{lipunov1996b} Lipunov V.M., Postnov K.A., Prokhorov M.E., 1996b, ed. Sunyaev R.A., The Scenario Machine: Binary Star Population
Synthesis, Astrophysics and Space Physics Reviews, vol. 9, Harwood
academic publishers

\bibitem[Lipunov et al. (1996c)]{lipunov1996c} Lipunov V.M., Postnov K.A., Prokhorov M.E.,
1996, A\&A, 310, 489

\bibitem[Lipunov et al. (1997)]{lipunov1997a} Lipunov V.M., Postnov K.A., Prokhorov M.E., 1997, MNRAS, 288, 245

\bibitem[Lipunov \& Shakura (1976)]{lipunov1976a} Lipunov V.M., Shakura
N.I., 1976, Soviet Astronomy Letters, v. 2, no. 4, p. 133

\bibitem[Lyne et al. (2004)]{lyne2004a} Lyne A.G. et al., 2004, Science, v. 303, pp. 1153-1157

\bibitem[Mardling (1995a)]{mardling1995a} Mardling R.A., 1995a,
ApJ, 450, 722

\bibitem[Mardling (1995b)]{mardling1995b} Mardling R.A., 1995b,
ApJ, 450, 732

\bibitem[McCrea (1953)]{mccrea1953a}  McCrea W.H., 1953, MNRAS, 113, 162

\bibitem[Mestel (1952)]{mestel1952a} Mestel L., 1952, 136, 583

\bibitem[Mitrofanov et al. (1977)]{mitrofanov1977a}  Mitrofanov I.G., Pavlov G.G., Gnedin Yu.N., 1977, Astron. Tsirk., 948, 5

\bibitem[Nieuwenhuijzen \& de Jager (1990)]{jager1990a} Nieuwenhuijzen H., de Jager
C., 1990, A\&A, v. 231, p. 134

\bibitem[Nomoto \& Kondo (1991)]{nomoto1991a} Nomoto K., Kondo Y., 1991, ApJ, v. 367, p. L19

\bibitem[Pacini \& Shapiro (1975)]{pacini1975a} Pacini F., Shapiro S.L., 1975, Nature, 255, 618

\bibitem[Paczyn'ski (1971)]{paczynski1971a} Paczyn'ski B., Annual Review of Astronomy and Astrophysics, 1971, v. 9, p. 183

\bibitem[Paczynski (1976)]{paczynski1976a} Paczynski B., 1976, IAU Proceedings of the Symposium no. 73,
Structure and Evolution of Close Binary Systems, Edited by P.
Eggleton, S. Mitton, and J. Whelan, Dordrecht, D. Reidel
Publishing Co., p. 75

\bibitem[Paczynski \& Sienkiewicz (1983)]{paczynski1983a} Paczynski B.,
Sienkiewicz R., 1983, ApJ, 268, 825

\bibitem[van Paradijs (1997)]{paradijs1997a} van Paradijs J., van den Heuvel E.P.J., Kouveliotou C., Fishman G.J., Finger M.H., Lewin
W.H.G., A\&A, v. 317, p. L9

\bibitem[Peters (1964)]{peters1964a} Peters P.C., Phys. Rev., 136,
1224

\bibitem[Peters \& Mathews (1963)]{peters1963a} Peters P.C.,
Mathews J., 1963, Phys. Rev., 131, 435

\bibitem[Pols \& Marinus (1994)]{pols1994a} Pols O.R., Marinus M.,
1994, A\&A, 288, 475

\bibitem[Press \& Teukolsky (1977)]{press1977a} Press W.H.,
Teukolsky S.A., 1977, ApJ, 213, 183

\bibitem[Pringle \& Rees (1972)]{pringle1972a} Pringle J.E., Rees M.J., 1972, A\&A, 21, 1

\bibitem[Prokhorov (1987)]{prokhorov1987a} Prokhorov M.E., 1987, Astron Tsirc., 1502, 1

\bibitem[Rappaport et al. (1982)]{rappaport1982a} Rappaport S., Joss P.C., Webbink
R.F., 1982, ApJ, v. 254, p. 616

\bibitem[Ritossa et al. (1996)]{ritossa1996a} Ritossa C., Garcia-Berro E., Iben
I., 1996, ApJ, v. 460, p. 489

\bibitem[Rogers \& Iglesias (1991)]{rogers1991a} Rogers F.J., Iglesias
C.A., 1991, Bulletin of the American Astronomical Society, v. 23,
p. 1382

\bibitem[Savonije \& van den Heuvel (1977)]{heuvel1977a} Savonije G.J., van den Heuvel E.P.J., 1977, ApJ, 214, L19

\bibitem[Schaller et al. (1992)]{schaller1992a} Schaller G., Schaerer D., Meynet G., Maeder
A., 1992, A\&AS, v. 96, p. 269

\bibitem[Shakura (1975)]{shakura1975a} Shakura N.I., 1975, Soviet Astronomy Letters, v. 1, no. 6, p. 223

\bibitem[Shakura \& Sunyaev (1973)]{shakura1973a} Shakura N.I., Sunyaev R.A., 1973, A\&A, 24, 337

\bibitem[Shvartsman (1970)]{shvartsman1970a} Shvartsman V.F., Soviet Astronomy, v. 14, p. 527

\bibitem[Shvartsman (1971a)]{shvartsman1971a} Shvartsman V.G., Soviet Astronomy, v. 14, p. 662

\bibitem[Shvartsman (1971b)]{shvartsman1971b} Shvartsman V.F.,
1971, Soviet Astronomy, v. 15, p. 342

\bibitem[Skumanich (1972)]{skumanich1972a} Skumanich A., 1972,
ApJ, v. 171, p. 565

\bibitem[Stothers \& Chin (1991)]{stothers1991a} Stothers R.B., Chin
C.-W., 1991, ApJ, v. 381, p. L67

\bibitem[Thorsett \& Chakrabarty (1999)]{thorsett1999a} Thorsett S.E., Chakrabarty D., 1999, ApJ, 512, 288

\bibitem[Tout \& Pringle (1992)]{tout1992a} Tout C.A., Pringle J.E., 1992,
MNRAS, 256, 269

\bibitem[Trimble (1983)]{trimble1983a} Trimble V., 1983, v. 303, p. 137

\bibitem[Tutukov \& Yungelson (1973)]{tutukov1973a} Tutukov A., Yungelson
L., 1973, Nauchnye Informatsii, v. 27, p. 70

\bibitem[Vanbevern et al. (1998)]{vanbeveren1998a} Vanbeveren D., de Donder E., van Bever J., van Rensbergen W., de Loore
C., 1998, New Astronomy, v. 3, p. 443

\bibitem[Van Bever \& Vanbeveren (2000)]{vanbever2000} Van Bever
J., Vanbeveren D., 2000, A\&A, 358, 462

\bibitem[van den Heuvel (1994)]{heuvel1994a} van den Heuvel E.P.J., in Shore S.N., Livio M., van den Heuvel E.P.J., 1994, Interacting Binaries, Springer-Verlag, p. 103

\bibitem[Varshavskii \& Tutukov (1975)]{varshavskii1975a} Varshavskii, V. I., Tutukov, A. V., 1975, SvA, 19, 142

\bibitem[Verbunt (1984)]{verbunt1984a} Verbunt F., 1984, MNRAS,
209, 227

\bibitem[Verbunt \& Zwaan (1981)]{verbunt1981a} Verbunt F., Zwaan
C., 1983, A\&A, v. 100, p. L7

\bibitem[Webbink (1979)]{webbink1979a} Webbink R.F., 1979,
Proceedings of the Fourth Annual Workshop on Novae, Dwarf Novae
and Other Cataclysmic Variables, Rochester N.Y., University of
Rochester, p. 426

\bibitem[Webbink (1985)]{webbink1985a} Webbink R.F., 1985, Stellar evolution and binaries, in Interacting Binary Stars,
Ed. Pringle J.E. and Wade R.A., Cambridge Astrophysics Series,
Cambridge University Press, p.39

\bibitem[Wickramasinghe \& Whelan (1975)]{wickram1975a} Wickramasinghe D.T., Whelan J.A.J., 1975, Nature, v. 258, p. 502

\bibitem[Weisberg \& Taylor (2003)]{weisberg2003a} Weisberg J.M., Taylor J.H., 2003, proceedings of "Radio Pulsars,"
Chania, Crete, August, 2002, ASP. Conf. Series, 2003, Edited by M.
Bailes, D.J. Nice, S.E. Thorsett

\bibitem[Woosley et al. (2002)]{woosley2002a} Woosley S.E., Heger A., Weaver T.A., 2002, Rev. Mod.
Phys., v. 74, p. 1015

\bibitem[Wex et al. (2000)]{wex2000a} Wex N., Kalogera V., Kramer M., 2000, \\ ApJ, 528, 401

\bibitem[Zangrilli et al. (1997)]{zangrilli1997a} Zangrilli L.,
Tout C.A., Bianchini A., 1997, \\ MNRAS, 289, 59

\bibitem[Zahn (1975)]{zahn1975a} Zahn J.-P., 1975, A\&A, 41, 329

\bibitem[Zahn (1989b)]{zahn1989a} Zahn J.-P., 1989, A\&A, 220, 112

\bibitem[Zahn \& Bouchet (1989b)]{zahn1989b} Zahn J.-P., Bouchet
L., 1989, A\&A, 223, 112

\bibitem[Zeldovich et al. (1972)]{zeldovich1972a} Zeldovich Ya.B., Ivanova L.N., Nadezhin D.K.,
1972, Soviet Astronomy, 16, 209

\end{thebibliography}
\end{document}